\documentclass[conference]{IEEEtran}
\IEEEoverridecommandlockouts
\usepackage[T1]{fontenc}
\usepackage{footnote}   
\usepackage{amsmath,amsfonts,amssymb}
\usepackage{comment}
\usepackage{lipsum}
\newtheorem{theorem}{Theorem}
\newtheorem{lemma}[theorem] {Lemma}
\usepackage{algorithm}% http://ctan.org/pkg/algorithms
\usepackage{algpseudocode}% http://ctan.org/pkg/algorithmic
\usepackage{graphicx}
\usepackage{textcomp}
\usepackage{mathtools}
\usepackage{caption}
\usepackage{float}
\usepackage{multirow}
\usepackage{subcaption}
\usepackage{xcolor}
\usepackage{tabularx}
\usepackage{bbm}
\newcommand{\taub}{\bar{\tau}}

\newcommand{\tauk}{{\tau}_{k}}
\newcommand{\tautil}{\tilde{\tau}}
\newcommand{\taustar}{\tau^{\ast}}

\newcommand{\Lotau}{L_0(\tau)}
\newcommand{\Lot}{L_0(0)}
\newcommand{\bbE}{\mathbb{E}}
\newcommand{\fng}{g(C_h)}

\usepackage{enumerate}
\newtheorem{remark}{Remark}
\newtheorem{proposition}{Proposition}
\newcommand{\chthetabybeta}{\frac{C_h-\theta}{\beta}}
\newcommand{\hzero}{h(0,0,0)}
\newcommand{\hzeroone}{h(0,0,1)}

\newcommand{\hqzeroone}{h(Q,0,1)}
\newcommand{\htauzero}{h(0,\tau,1,0)}
\newcommand{\htauone}{h(0,\tau,1,1)}
\newcommand{\Qhat}{\hat{Q}}
\newcommand{\Qbar}{\bar{Q}}
\newcommand{\Qstar}{Q^{\ast}}
\newcommand{\remove}[1]{}
\renewcommand{\baselinestretch}{0.95}
\def\BibTeX{{\rm B\kern-.05em{\sc i\kern-.025em b}\kern-.08em
    T\kern-.1667em\lower.7ex\hbox{E}\kern-.125emX}}
\begin{document}

\title{Dynamic Content Caching with Waiting Costs via Restless Multi-Armed Bandits 
{\footnotesize \textsuperscript{}}
%\thanks{This work was supported jointly by Centre for Network Intelligence, Indian Institute of Science (IISc), a CISCO CSR initiative and Aircel TCoE project 39010C.}
}

\author{\IEEEauthorblockN{ Ankita Koley, Chandramani Singh}
\IEEEauthorblockA{\textit{Department of Electronic Systems Engineering} \\
\textit{Indian Institute of Science}\\
Bangalore 560012, India \\
Email:\{ankitakoley, chandra\}iisc.ac.in}}

\maketitle


%\email{webmaster@marysville-ohio.com}
%\affiliation{%
 % \institution{Institute for Clarity in Documentation}
  %\streetaddress{P.O. Box 1212}
  %\country{USA}


\begin{abstract}
We consider a system with a local cache  connected to a backend server and an end user population. A set of contents are stored at the the server where they continuously get updated. The local cache keeps copies, potentially stale, of a subset of the contents. The users make content requests to the local cache  which  either can serve the local version if available or can fetch a fresh version or can wait for additional requests before fetching and serving a fresh version. Serving a stale version of a content incurs an age-of-version~(AoV) dependent ageing cost, fetching it from the server incurs a fetching cost, and making a request wait incurs a per unit time waiting  cost. We focus on the optimal actions  subject to the cache capacity constraint at each decision epoch, aiming at minimizing the long term  average cost. We pose the problem as a Restless Multi-armed Bandit~(RMAB) Problem and propose a Whittle index based policy which is known to be asymptotically optimal. We explicitly characterize the Whittle indices. We numerically evaluate the proposed policy and also compare it to a greedy policy. We show that it is close to the optimal policy and substantially outperforms the  exising policies. 
\end{abstract}

%\ccsdesc[500]{Do Not Use This Code~Generate the Correct Terms for Your Paper}
%\ccsdesc[300]{Do Not Use This Code~Generate the Correct Terms for Your Paper}
%\ccsdesc{Do Not Use This Code~Generate the Correct Terms for Your Paper}
%\ccsdesc[100]{Do Not Use This Code~Generate the Correct Terms for Your Paper}


 % \includegraphics[width=\textwidth]{sampleteaser}
  %\caption{Seattle Mariners at Spring Training, 2010.}
  %\Description{Enjoying the baseball game from the third-base
  %seats. Ichiro Suzuki preparing to bat.}
  %\label{fig:teaser}
\maketitle

\section{Introduction}
Real-time applications, such as social media platforms, e-commerce sites, and news websites, are the hotspots of the internet. Users widely use these real-time applications, so timely delivery of contents is essential for quality of service. These contents  are highly dynamic and can quickly become outdated. The recent versions of the contents can be directly delivered from the backend servers. However, due to the massive demands for these applications, enormous communication resources between the servers and the  users are warranted and it also makes  timely dissemination of contents challenging. In particular, directly delivering content from the servers can lead to network congestion. Alternatively, contents can be stored at the local caches deployed by the Content Distributing Networks. However, it is essential to periodically replace the cached content to ensure that caches always have popular contents and the users receive their fresh versions. So, dynamic content caching plays a vital role in  ensuring resource-efficient and timely delivery of contents.

Age of information~(AoI), which is defined as time elapsed since the last fresh version is fetched from the backend server has been widely used as a metric for content freshness~\cite{kaul2012status,FAHIM2024102415} . However, AoI  does not truly reflect the relevance of a content's copy, which is a function of its   freshness and depends on the frequency of the content updates at the backend server. For instance, a high AoI does not diminish a content's value if it has not been updated at the  server. The age of version~(AoV) is a newly introduced metric that captures the number of updates at server since the content is fetched to the local cache~\cite{9488731}. It more accurately measures  freshness of the content. However, the AoV is of a content is not observable at the cache until content is fetched again, which make the AoV optimization hard.

In general, a local cache must determine which contents to retain and when to replace them with fresh versions. This task is challenging due to the limited capacity of the cache, the stochastic nature of content updates, their popularities and request arrivals, and most importantly, the lack of access to the AoVs of the contents at the cache.  

In this work, we consider a  backend server and a connected finite capacity local cache which  caches a subset of the contents and serves a population of the end users. The contents dynamically get updated at the server. Upon receiving a request for a content, the cache can serve its locally cached version, can fetch it from the  server and serve, or can wait for additional requests for it before fetching and serving it. So, in addition to the content fetching costs, the cache incurs ageing costs on serving stale versions of the contents and waiting costs due to queuing of requests. We aim to minimize the average cost subject to the cache capacity constraints. We pose this problem as a Restless Multi-armed Bandit~(RMAB) problem treating each content as an arm. Contents' states evolve as a Partially Observed Markov Decision Process~(POMDP) that suffers from curse of dimensionality. We propose a Whittle index-based policy that performs very close to the optimal policy and is provably asymptotically optimal.

\subsection{Related Work}
Caching of dynamic contents has gained a lot of attention recently. Here,  we  discuss a few closely related works on dynamic content caching. Abolhassani et al.~\cite{9155324} study  content freshness in single and multi user scenarios where the requested content is served from the cache if there is a cache hit and fetched from the server otherwise. Additionally, the local cache keeps updating the cached contents through cache checks that constitute a Poisson process. In a latter work~\cite{9488731}, they consider a more flexible strategy that fetches a fresh version of  a requested content from the server even when there is a cache hit, depending upon the cached version's AoV. They propose a static policy and show that it is asymptotically optimal. We propose a Whittle index based dynamic policy in ~\cite{10723415} which outperforms the one proposed in~\cite{9488731}.
More recent works have extended the study of~\cite{9488731,9155324} to incorporate wireless links~\cite{9771060}, multiple caches~\cite{10045757} and push based algorithms~\cite{10452408}.
\remove{
 We consider a more flexible scenario where the local cache can {\it wait} for additional requests before serving the requested content. This leads to queueing of requests waiting at the local cache for each content. In addition to ageing and fetching costs, we also consider per unit waiting cost for each request, and we aim to minimize the sum of average waiting cost, fetching cost, and ageing cost subject to cache capacity constraints. 
 }

We frame the content caching and delivery problem subject to the cache capacity constraint as  a 
restless multi-armed bandit (RMAB) where each content represents an arm. In general, the RMABs admit index based policies and a subclass of these policies, the Whittle index-based policies, are known to be  asymptotically optimal. In general Whittle indices are hard to compute. Several recent works have proposed efficient algorithms to numerically compute Whittle indices~\cite{Akbarzadeh_Mahajan_2022,gast2023testing}. 
Very few works have offered explicit characterizations of Whittle indices for their RMAB problems~\cite{glazebrook2005index,5605371,tripathi2017age} 
\remove{In general, the RMABs are intractable, and the Whittle index-based policy usually performs very close to the optimal policy if the content is indexable. \remove{ Obtaining the Whittle index based policy requires solving of the single content problem first. The single content problem evolves as a Markov Decision Process (MDP) where the local cache is aware of the update rate of the each content but not the exact time of the updates.}
}
Finally, the existing works on RMABs~(e.g.,~\cite{9165970,tripathi2017age,5605371,10260280,10570315}) are limited to scenarios where the action space consists of only two actions, pulling or not pulling the arms. However, our setup allows more choices at each decision epoch~(see Figure~\ref{fig:finite_capacity_with_queue}) which makes the analysis more challenging.

\subsection{Our Contribution}
We study a content fetching, caching and delivery problem in a system with a finite capacity local cache. 
To the best of our knowledge, this is the first work to consider both content ageing~(AoV) and waiting costs. We see that allowing {\it wait} action for the requested contents reduces the overall average cost. Following is a preview of our contribution.
\begin{enumerate}
     \item We first analyze the optimal content caching and delivery problem assuming infinite cache capacity and provide explicit optimal policy. 
    \item For the finite capacity case, we formulate the problem as a RMAB,
    show its indexability, and design a Whittle index based policy. We provide explicit expressions for the Whittle indices. 
    \item We also demonstrate that our policy performs close to the optimal policy. Since there are no prior works, we compare our policy with a greedy policy and show that our policy yields much lower cost than the greedy policy.   
\end{enumerate}

\section{System Model}
\label{sec:sys_model_queue}We consider system with a local cache connected to a backend sever connected via the core network and an end user population. The server hosts a set of dynamic contents $[N] \coloneq \{1,2,\cdots,N\}$. The local cache  can store up to $M$ contents where $M < N$. The contents are dynamic and get updated at the server. At any time, one or more contents in the local cache may be stale copies. The users send requests to the local cache for specific contents. On being requested a content, the local cache serve its locally cached, potentially stale, copy if available, can fetch a fresh copy from the server and serve, or can choose to wait for additional requests for the same content before fetching a fresh copy from the server and serving. Furthermore, after fetching the content from the server, the local cache may cache it and evict an existing content. 

Let $\mathcal{E}(t) \subset [N]$ denote the set of cached contents at the local cache at time $t$. Note that $\mathcal{E}(t)$ can potentially change only  on the content request  epochs.

\paragraph*{Content updates} The contents get updated according to independent Poisson processes, with $\lambda_n$ being the update rate for content $n$. The server always holds the most recent versions of these contents.
Let $V^n(t) \in \mathbb{Z}_+$ be the number of times content $n \in \mathcal{E}(t)$ has been updated since it was last fetched by the local cache as of time $t$.  We refer to $V^n(t)$ as the age-of-version~(AoV) of content $n$ at time $t$. Note that AoVs of the cached contents are not observable at the local cache.

\paragraph*{Content requests} The aggregate request process of the end users is a Poisson process with rate $\beta$. Each request could be for the $n^{th}$ content with probability $p_n$ independently of the other requests. \footnote{We use the phrases ``$n^{th}$ content'' and ``content $n$'' interchangeably.}Here $p_n, 1 \leq n \leq N$ denote the relative popularities of the contents and $\sum_n p_n = 1$. For instance, the content popularity on World Wide Web is widely modelled as {\em Zipf's distribution} 
wherein $p_n \propto 1/n^{\alpha}$ for the $n^{th}$ most popular content~\cite{Breslau99}. Under the proposed request dynamics, the $n^{th}$ content's request arrivals constitute a Poisson process with rate $p_n\beta$. 
  
\paragraph*{Request queues} Let $Q^n(t)$ be the number of requests for the $n^{th}$ content waiting for service at the local cache. Note that $Q^n(t)$ stays at zero as long as the local cache keeps serving the cached version of the $n^{th}$ content. However, it grows if the local cache chooses to fetch and serve a fresh copy of the content but waits for additional requests for it before doing so. More specifically, the request queue evolves as follows. On arrival of a request for the $n^{th}$ content if the local cache serves the cached version or fetches a fresh copy and serves then $Q^n(t)$ becomes $0$. If the local cache waits for additional requests before serving the tagged request then $Q^n(t)$ increases by $1$.  
\remove{
\begin{enumerate}
    \item On arrival of a request for the $n^{th}$ content if the local cache serves the cached version or fetches a fresh copy and serves then $Q^n(t)$ becomes $0$.
    \item If the local cache waits for additional requests then $Q^n(t)$ increases by $1$. 
\end{enumerate}
}
\paragraph*{Content fetching,  ageing and waiting costs}
If a content, say content $r$, is requested at time $t$, one of the following scenarios may be encountered. 

\begin{enumerate}
\item {\it Content $r$ is cached:} In this case,
 \begin{enumerate}
     \item the local cache may serve the cached version incurring an {\it ageing cost} $c_a V^r(t)$ where $c_a$ is the ageing cost per unit AoV. If the cache serves the cached version to the $Q^r(t)$ waiting requests, it incurs  an {\it ageing cost} $c_a V^r(t)$ for each of those. The ageing costs reflect the users' aversion to stale contents.  
     \item the cache may fetch and serve the fresh version of content $r$ to the request arrived at $t$ and also to the $Q^r(t)$ waiting requests. This action incurs a {\it fetching cost} $c_f$ which is the cost of procuring the content from the server. 
     \item it may wait for additional requests before serving the requests for content $r$.
 \end{enumerate}
\item {\it Content $r$ is not cached:} In this scenario, the local cache may fetch and serve the latest version of content $r$, incurring a {\it fetching cost} $c_f$,   or may wait for additional requests for content $r$ before fetching and serving. 
 \end{enumerate}

\begin{figure}[h]
    \centering
    \includegraphics[width=\linewidth]{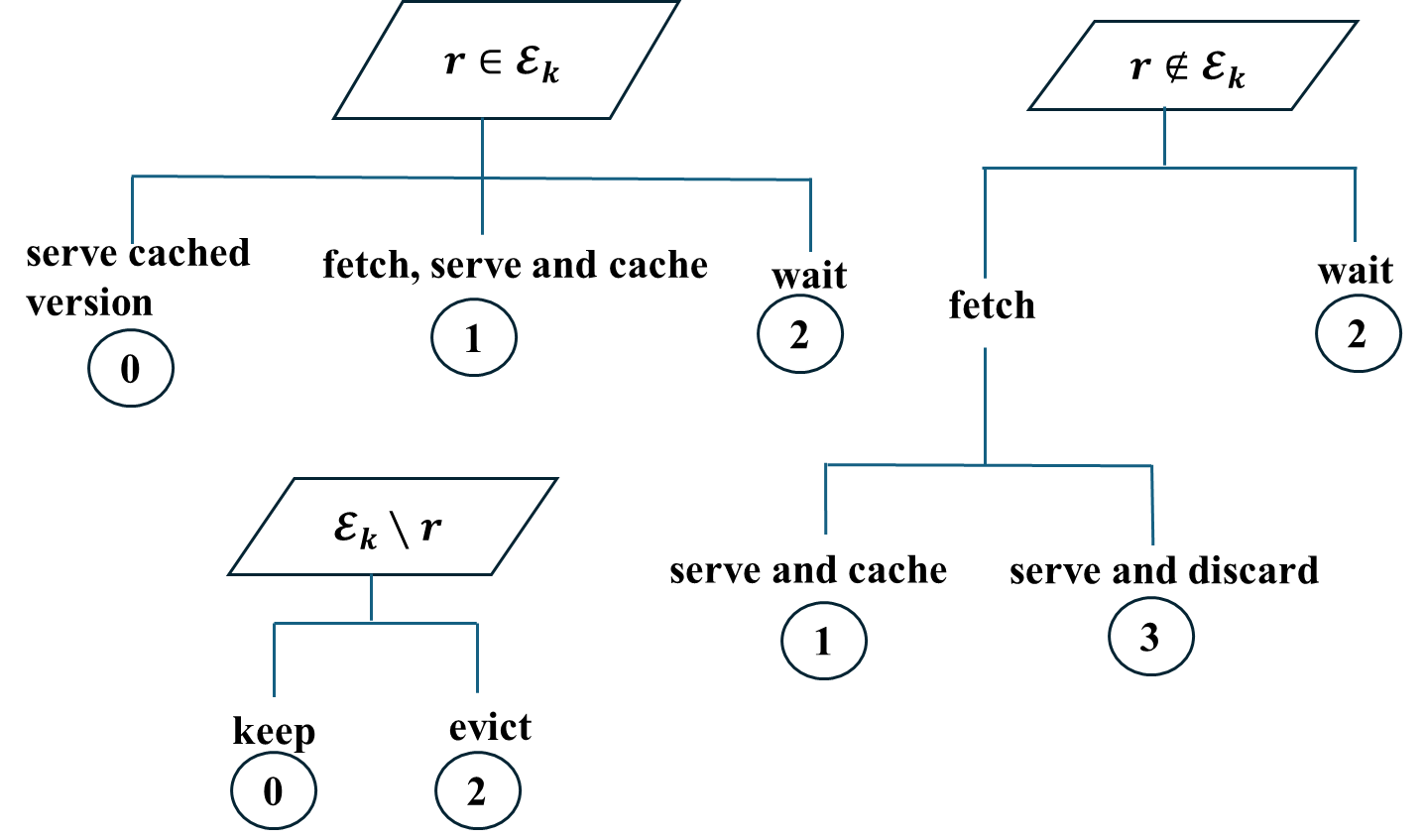}
    \caption{Possible actions for the contents $\{r\}\cup\{\mathcal{E}_k\}$. For instance, Action~$2$ is applicable for all the contents in $\{r\}\cup\{\mathcal{E}_k\}$, however, it means {\it wait} for  content $r$ and {\it evict} for the contents in $\mathcal{E}_k\setminus \{r\}$.}
    \label{fig:finite_capacity_with_queue}
\end{figure}

Each waiting request also has an associated waiting cost $c_w$ per unit time. So, the local cache also incurs a running waiting cost $(\sum_n Q^n(t))c_w$ per unit time.
The waiting costs quantify the users’ quality of experience. The fetching, ageing and waiting costs can be different for different contents. 
Let us represent various actions using numbers $0,1,2$ and $3$ whose annotations are as in Figure~\ref{fig:finite_capacity_with_queue}. We let the local cache serve only content $r$ at $t$ although there could be pending requests for contents $\mathcal{E}(t) \setminus \{r\}$. It can be argued that if
the optimal action for a content request on its arrival is to wait for more requests for it, then the optimal action for it continues to be wait until a new request for it arrives. For brevity we omit the proof and do not consider ``serve'' action for the contents $\mathcal{E}(t) \setminus \{r\}$.  Clearly, not all actions are available for all the contents in ${\cal E}(t) \cup \{r\}$. Using $a^n$ to denote the action vis a vis content $n \in \mathcal{E}(t) \cup \{r\}$; $a^r \in \{0,1,2\}$ if $r \in \mathcal{E}(t)$,  $a^r \in \{1,2,3\}$ if $r \notin \mathcal{E}(t)$, and $a^n \in \{0,2\}$ for $n \in \mathcal{E}(t) \setminus \{r\}$. 

We aim to develop a policy that prescribes actions on request arrival epochs to minimize the long term average content fetching, ageing, and waiting costs. 

 \subsection{The Optimal content Caching and  Delivery Problem}
 \label{sec:optimal-problem}
 Let $t_k$ and $R_k$ denote the $k^{th}$ request epoch and the requested content at $t_k$, respectively. Let $V^n_k$ and $a^n_k$ denote the AoV of content $n$ at $t_k$ and the action vis a vis content $n$ at $t_k$, respectively, for $n \in \mathcal{E}_k \cup \{R_k\}$. For each $n \in [N]$, let $Q^{n}_k$ be the request queue length of content $n$ immediately prior to the arrival of request $R_k$.  

 We now focus on the cost incurred by the local cache in $[t_k, t_{k+1})$. Recall that the local cache incurs instantaneous fetching and ageing costs for the requested content $R_k$ at $t_k$, and also the running waiting cost for all the contents $n\in N$ for which $Q^n_k>0$.  Let $C^n_k$  be the cost associated with content $n$.
 Then
 \begin{align*}
     C^{R_k}_k = &\mathbbm{1}_{\{a_k^{R_k} \in \{1,3\}\}} c_f+(Q^{R_k}_k+1)\mathbbm{1}_{\{a_k^{R_k} =  0\}} c_a V^{R_k}_k\\
     & +(Q^{R_k}_k+1)\mathbbm{1}_{\{a_k^{R_k} =  2\}}c_w(t_{k+1}-t_k)
\end{align*} 
and for all $n \in [N]\setminus R_k$,
\[
C^n_k=Q^{n}_kc_w(t_{k+1}-t_k).
\]
So, the local cache incurs the total cost $ \sum_{n\in [N]}C^n_k$ in $[t_k, t_{k+1})$. Let $M_k^n \in \{0,1\}$ denote the caching status of content $n$; it is $1$ if and only if content $n$ is in the cache during $[t_k,t_{t+1})$. Clearly, 
\[M_k^n =
\begin{cases}
    \mathbbm{1}_{\{a^n_k\in\{0,1\}\}}+ \mathbbm{1}_{\{a^n_k=2,n\in\mathcal{E}_k\}} \text{ if } n=R_k,\\
    \mathbbm{1}_{\{ a_k^n=0\}} \text{ if } n\in\mathcal{E}_k\setminus R_k,\\
    0, \text{ otherwise.}
\end{cases}
\]
 Further, let $A(T)$ denote the total number of requests until time $T$; $A(T) = \max\{k: t_k \leq T\}$ and $\mathbb{E}[A(T)] = \beta T$.
 We aim to minimize the average cost which can be formally expressed as follows. 
 \begin{align}
\min & \lim_{T \to \infty}\frac{1}{T}\mathbb{E}\left[\sum_{k=1}^{A(T)}\sum_{n\in [N]}C^n_k\right]\label{eq:POMDP_with_finite_cache}\\
\text{subject to }& {\sum_{n\in [N]}} M^n_k= M, \,\forall k \label{constraint_POMDP_with_finite_cache}
\end{align}

Treating $({\cal E}_k,R_k,Q^n_k, V^n_k,n \in {\cal E}_k), (a^n_k,n \in {\cal E}_k \cup \{R_k\})$ and $(R_{k+1},V^n_{k+1},n \in {\cal E}_{k+1})$ as the state, the action and the random noise, respectively, at $t_k$, we can pose the above problem as a MDP.
Recall that $V^n_k$'s are not observable to the local cache at $t_k$ prior to taking any action. 
So, we have a POMDP at hand.
This POMDP suffers from the curse of dimensionality. However, we notice that the actions and state evolutions of different contents are only coupled through the cache capacity constraint~\eqref{constraint_POMDP_with_finite_cache}. So, we  formulate the problem as a RMAB problem where each content represents an arm. We show that the RMAB problem is indexable and propose a Whittle index based policy~\cite{whittle1988restless}. We also  explicitly characterize the Whittle indices. Observe that, unlike the classical RMAB setting with binary actions, we have more state dependent actions, rendering the  problem complex.\footnote{The difference between this work and~\cite{10723415} is the additional \it{wait} action. Allowing \it{wait} option adds $N$ more dimensions, the queue lengths, to the system’s state. It also substantially changes the state dynamics, the analysis and the caching policy.}
Before describing the RMAB formulation of content caching and delivery problem in Section~\ref{sec:finite_cache_capacity_with_waiting}, we discuss  it in the special case of infinite cache capacity. 
%\input{introduction}
%\input{system_model}
%\input{RMAB_formulation}
 %\section{Single content with infinite caching capacity with missing cost and wireless link between sever and cache}
 %\input{single content with infinite cache}
 \section{Infinite Caching Capacity}\label{sec:inf_cap_queue}
\remove{
We consider a local cache associated with a local cache, which in turn connected to the central server via wired link and the user requests a content from the local cache according to Poisson process with rate $\beta$. The content gets updated at the central server according Poisson process with rate $\lambda$. }\remove{ A version of the content is always cached at the local cache and let $V(t)$ be the number of updates since it was last fetched until time $t$. $V(t)$ is the {\it age-of-version} (AoV) of the content at time $t$. 
}

We assume that the local cache has an infinite capacity cache, i.e., $M= \infty$. In this case, the actions for different contents need not depend on each other. So, we can decouple  the optimal caching and content delivery problem into $N$ separate problems, one corresponding to each content.  

Let us now focus on the the optimal caching and content delivery problem associated with a tagged content, Say content $n$,  Here, we let  $t_k$ denote the $k^{th}$ request epoch for content $n$.
For brevity, in this section we omit the superscripts representing the content indices in the variables. 
For instance, we let $Q_k$ and $V_k$ be the number of pending requests and the AoV of  content $n$, respectively, at $t_k$. Let $a_k$ denote the action taken by the local cache at $t_k$;  $a_k \in \mathcal{A}:=\{0,1,2\}$, where $a_k=0$ denotes {\it serve the cached copy}, $a_k=1$ denotes {\it fetch and serve}  and $a_k=2$ denotes {\it wait}.

 Let $C_k$ denote the cost incurred by the local cache vis a vis content $n$ at $t_k$. Then 
\[C_k=\begin{cases}
    c_aV_k(Q_k+1) \text{ if }a_k=0\\
    (Q_k+1)c_w(t_{k+1}-t_k) \text{ if }a_k=2 \\
    c_f \text{ if }a_k=1
\end{cases}\]

We aim to minimize the average cost, i.e., to solve  
\begin{align}
\min & \lim_{T \to \infty}\frac{1}{T}\mathbb{E}\left[\sum_{k=1}^{A(T)}C_k\right]\label{eq:POMDP_with_infinite_cache}
\end{align}

We can again pose the problem as a POMDP with decision epochs $t_k$, states $(Q_k,V_k)$, actions $a_k$ and noise $V_{k+1}$. 
 In the following, we reformulate the problem as an MDP and find the optimal policy. 

\subsection{Problem Reformulation}
The local cache does not know of the updates of content $n$ at the server until it fetched the content. In particular, the local cache does not know $V_k$ prior to taking an action at $t_k$.   
Nevertheless, it knows the time since last fetch of content $n$ at $t_k$, which we denote by $\tau_k$; $\tauk=t_k-\max\{l_k:a_k=1\}$. So, the local cache can compute the expected ageing  cost at $t_k$, $\bbE[V_k]=\lambda\tauk$. Moreover, $\tau_{k+1}$ is independent of $(Q_j,\tau_j), j \leq k-1$ given $(Q_k,\tau_k)$ and $a_k$. Hence, we can reformulate~\eqref{eq:POMDP_with_infinite_cache} as MDP considering $(Q_k, \tau_k)$ as the state at $t_k$.  Now, the state space is $\mathcal{S}:=\{ (Q,\tau), Q\in\mathbb{Z}_+ \text{ and } \tau\in\mathbb{R}_+ \}$ and the action space continues to be $\mathcal{A}$. 
Following are the state dynamics and the single stage costs for this for this MDP.

\paragraph*{State dynamics}
Given $(Q_k, \tau_k)=(Q,\tau)$, the state at $t_{k+1}$ will be
\begin{align}
(Q_{k+1},\tau_{k+1})=\begin{cases}
        (0,\tauk+\Delta \tau) \text{ if } a_k=0\\
        (Q+1,\tauk+\Delta \tau) \text{ if } a_k=2\\
        (0,\Delta \tau) \text{ if } a_k=1
    \end{cases}
\end{align}
where $\Delta \tau~\sim$ Exponential$(\beta)$.
\paragraph*{Cost}
The expected value of the single stage cost will be 
\[c(Q,\tau,a)=\begin{cases}
      c_a\lambda \tau(Q+1) \text{ if }a=0,\\
    \frac{c_w}{\beta}(Q+1) \text{ if }a=2, \\
    c_f \text{ if }a=1.
\end{cases}\]
We define a stationary policy $\pi$ to be a mapping from the state space to action space. Let $\Pi$ be the set of all stationary policies. It can be shown that there exists a stationary optimal policy for the MDP at hand~\cite{bertsekas2011dynamic}.  Hence the MDP problem is 
\begin{equation}
\min_{\pi \in \Pi}J_{\pi}(Q,\tau) 
\end{equation}
\begin{align*}
&\text{where }\lefteqn{J_{\pi}(Q,\tau)} \\
&= \lim_{T \to \infty}\frac{1}{T}\mathbb{E}\left[\sum_{k=1}^{A(T)}c(Q_k,\tauk,\pi(Q_k,\tau_k))\vert(Q_0,\tau_0){=}(Q,\tau)\right].
\end{align*}

Since the embedded discrete time Markov chain $(Q_j,\tau_j), j \geq 0$ has a single recurrent class, $J_{\pi}(Q,\tau)$ are independent of $(Q,\tau)$
~\cite[Section~4.2]{bertsekas2011dynamic}. Moreover, the Bellman's equations for the continuous time MDP can be written in the same way as for discrete-time problems. 
Let $h(Q,\tau)$ and $\theta$ be the relative cost function and the optimal cost, respectively. Then  Bellman's equations for different states are as follows~\cite[Section 5.3]{bertsekas2011dynamic}.
\begingroup
\allowdisplaybreaks
  \begin{align}
        h(Q,\tau){=}&\min\left\{-\frac{\theta}{\beta}{+}c_a\lambda\tau(Q+1){+}\int_{0}^{\infty}\beta e^{-\beta t}h(0,t{+}\tau)dt,\right.\nonumber\\
        &\left. -\frac{\theta}{\beta}+\frac{c_w}{\beta}(Q+1)+\int_{0}^{\infty}\beta e^{-\beta t}h(Q+1,t+\tau)dt, \right.\nonumber\\
        &\left. c_f-\frac{\theta}{\beta}+\int_{0}^{\infty}\beta e^{-\beta t}h(0,t)dt\right\}.
         \label{bellman_eqn_with_queue_infinite_capacity}
  \end{align}
  \endgroup
  The optimal policy is a threshold based policy with threshold on both the queue length and the time since last fetch. The following theorem provides this policy. 
  \begin{theorem}\label{thm:inf_cap_wait}
      The optimal policy $\pi^{\ast}$ is as follows.
      \[\pi^{\ast}(Q,\tau)=\begin{cases}
              0 \text{ for }\tau\leq \taustar, \\
              2 \text{ for }Q< Q^{\ast} \text{ and }\tau>\taustar,\\
              1 \text{ for }Q\geq Q^{\ast} 
          \end{cases}\]
          where $\taustar$ and $Q^{\ast}$ are the solutions to the following equations.
\begin{align*}
          &\tau=-\frac{(Q+1)}{\beta}+\frac{1}{\beta}\sqrt{(Q+1)^2+\frac{2\beta c_f}{c_a\lambda}+\frac{Q(Q+1)c_w}{c_a\lambda}}, \\
           &Q=\left\lfloor \frac{\beta c_a\lambda\tau}{c_w}\right\rfloor.    \end{align*}
           Moreover, the optimal cost $\theta=\beta c_a\lambda \taustar$.
  \end{theorem}
  \begin{IEEEproof}
    Please see Appendix~\ref{apndx:prf_inf_cap}.
  \end{IEEEproof}
 
 \section{Finite Caching Capacity}
\label{sec:finite_cache_capacity_with_waiting}
 In this section, we propose a Whittle index based policy for the optimal content caching and delivery problem with the cache constraints, i.e,~\eqref{eq:POMDP_with_finite_cache} subject to~\eqref{constraint_POMDP_with_finite_cache}.  We first consider a problem with the following relaxed constraint.
 \begin{align}
  \lim_{T\to \infty}\frac{1}{\beta T}\mathbb{E}\left[\sum_{k=1}^{A(T)}\sum_{n=1}^N M^n_k\right] = M. \label{relaxed_constraint_fnite_cache} 
\end{align} 
 The relaxed constraint~\eqref{relaxed_constraint_fnite_cache} applies to the expected total number of contents cached over the infinite horizon rather than at each epoch.  
To minimize the optimal caching problem~\eqref{eq:POMDP_with_finite_cache} subject to the relaxed constraint~\eqref{relaxed_constraint_fnite_cache}  we write the Lagrangian with multiplier   $C_h$ as follows.
\begin{align}
   &\lim_{T \to \infty}\frac{1}{T}\mathbb{E}\left[\sum_{k=1}^{A(T)}\sum_{n=1}^N C^n_k\right]+C_h\left( \lim_{T\to \infty}\frac{1}{\beta T}\mathbb{E}\left[\sum_{k=1}^{A(T)}\sum_{n=1}^N M^n_k\right]{-}M\right)\nonumber\\
    &=\sum_{n=1}^{N}\lim_{T \to \infty}\frac{1}{T}\mathbb{E}\left[\sum_{k=1}^{A(T)}C^n_k
    -\frac{C_h}{\beta}M^n_k \right]  -C_hM\nonumber\\
    & \eqcolon \sum_{n=1}^{N}V^n_{\pi}(C_h)-C_hM\eqcolon V(C_h). \label{eq:POMDP_relaxed_problem}
\end{align}

Minimizing~\eqref{eq:POMDP_with_finite_cache} subject to~\eqref{relaxed_constraint_fnite_cache} entails first minimizing~\eqref{eq:POMDP_relaxed_problem} for all $C_h$ and then maximizing the optimal values, say $V(C_h)$, over all $C_h$. The optimal policy for the relaxed problem need not be a feasible policy for the original problem as it need not satisfy the hard capacity constraints at each $t_k$. However, this policy provides ``Whittle indices'' associated with different states of all the  contents which in turn can be used to design a feasible policy  for the
original problem~\cite[Section~3.3.3]{zhao2022multi}. The Whittle index based policy is asymptotically optimal for the original problem as the number of contents $N$ approaches infinity and the cache capacity $M$ also grows proportionately~\cite[Conjecture~3.16]{zhao2022multi}.

Observe that in  minimizing~\eqref{eq:POMDP_relaxed_problem} the optimization problems corresponding to different contents are decoupled. Moreover, the Lagrange multiplier $C_h$ can be interpreted as the holding cost per unit time for the cached contents. So, in Section~\ref{sec:single-content-holding-cost}, we focus on a single content problem  with holding cost $C_h$. 
Note that the optimal policy will always keep the content if $C_h<0$, resulting in non-negative Whittle indices for all the contents in all the states.\remove{\footnote{We formally introduce Whittle indices in Section~\ref{eq:indexability and whittle index with waiting}.}} Hence, as in~\cite{aalto2019whittle}, we  consider $C_h\geq 0$.
We design a Whittle index based policy in Section~\ref{eq:indexability and whittle index with waiting}.

\subsection{Single content problem with holding cost}
\label{sec:single-content-holding-cost}

Suppose a content, say content $r$, is requested at  $t_k$ and $r \in \mathcal{E}_k$. Since there is a running cost $C_h$ for keeping any content in the cache, evicting  content $r$ after serving is also a potential action at $t_k$. Let $t_k$ be the $k^{th}$ request epoch across all the contents. As in section~\ref{sec:sys_model_queue}, let $R_k$ be the requested content at $t_k$ and  $Q_k$ and $V_k$ be the queue length and AoV, respectively, of content $r$ at $t_k$.  Let us also define a binary variable $b_k$ indicating whether the requested content at $t_k$ is  content $r$; $b_k = 1$ if $R_k = r$ and $0$ otherwise. Let $Y_k$ be another binary variable such that $Y_k=1$ if content $r$ is cached at $t_k$, i.e., if $r \in \mathcal{E}_k$. We can define $s_k \coloneqq (Q_k,\tauk, Y_k, b_k)$ if $Y_k=1$ and $s_k \coloneqq (Q_k,Y_k,b_k)$ if $Y_k=0$ to be the state of the system at $t_k$.
The next proposition shows  that we can restrict to policies in which {\it wait} action is never followed by {\it serve the cached version} action.   
\begin{proposition} \label{prop:policy}
   Let us consider a policy $\pi$ that takes the following actions for  some $\tau_2\ > \tau_1$ and $q_1$. 
      \[
\pi(q_1, \tau_1, 1, 1)= \text{wait and }\]
\[\pi(q_1+1, \tau_2, 1, 1)= \text{serve the cached version}.
\]
Let $\bar{\pi}$ be another policy 
 that takes identical actions as $\pi$ in all the states but $(q_1, \tau_1, 1, 1)$ and
\[
\bar{\pi}(q_1, \tau_1, 1, 1) = \text{serve the cached version}.
\] 
Then the average cost under $\bar{\pi}$ is at most the average cost under $\pi$. 
\end{proposition}
 
\begin{figure}[h]
    \centering
    \includegraphics[width=0.8\linewidth]{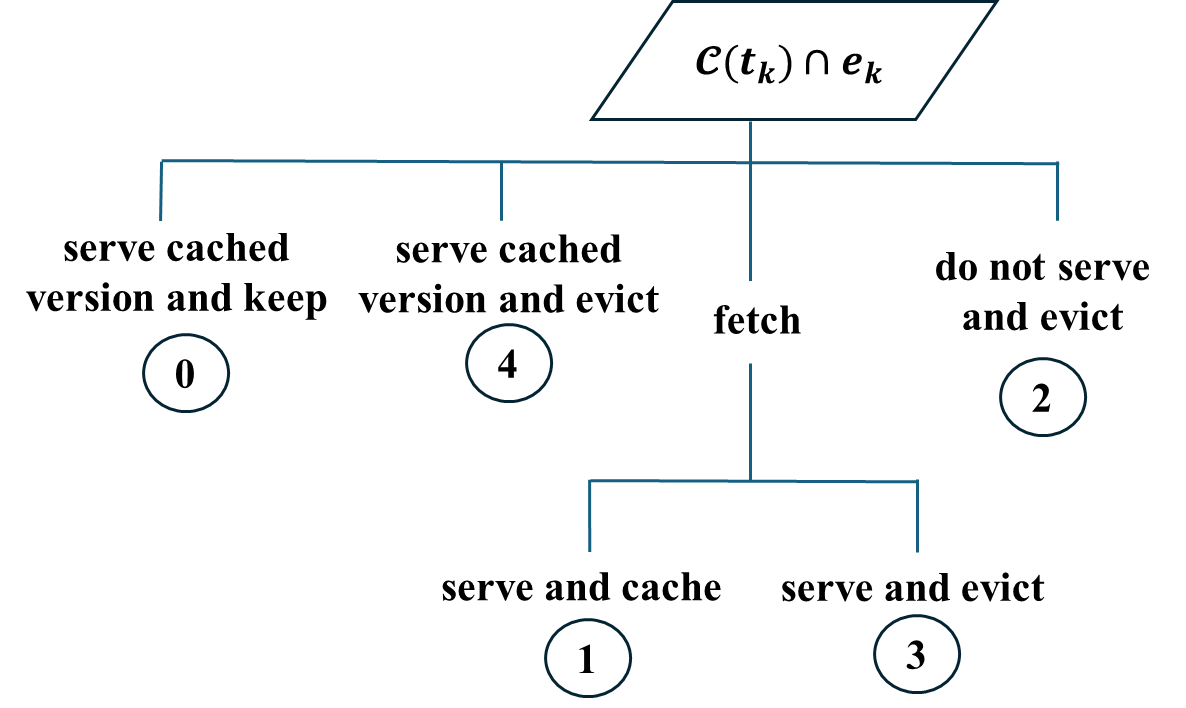}
    \caption{Possible actions for the tagged content with holding cost. Here we have additional actions, Action~$3$ denoting {\it fetch, serve and evict} and Action~$4$ denoting {\it serve the cached version and evict},  compared to Figure~\ref{fig:finite_capacity_with_queue}.}
    \label{fig:with_holding_cost_with_queue}
\end{figure}

\begin{IEEEproof}
Note that the system enters instantaneously state $(0,0)$ after every {\it fetch and serve} action. Therefore, the successive {\it fetch and serve} actions constitute renewal epochs. So, we only need to consider the average cost incurred in a renewal cycle. Further notice that the time since last fetch, $\tau$, monotonically increases in a cycle before resetting to zero at the end of the cycle. The states $(q_1, \tau_1, 1, 1)$ and $(q_1+1, \tau_2, 1, 1)$ occur in a renewal cycle if and only if there are $q_1$ requests for the tagged content are waiting to be served at $\tau_1$ and two successive requests arrive at $\tau_1$ and $\tau_2$. The policies $\pi$ and $\bar{\pi}$ incur identical expected cost in the renewal cycles that do not contain the states $(q_1, \tau_1, 1, 1)$ and $(q_1+1, \tau_2, 1, 1)$. So, let us focus on a renewal cycle containing these states. In such a cycle the costs incurred by $\pi$ and $\bar{\pi}$ differ only in the interval  $[\tau_1, \tau_2]$.
Let $C_{\pi}([\tau_1, \tau_2])$ and $C_{\bar{\pi}}([\tau_1, \tau_2])$ denote the expected cost given $\tau_1,\tau_2$ and $q_1$ under  $\pi$ and $\bar{\pi}$, respectively.
\remove{Note that, $q_2>q_1$ implies two possibilities $(a)$ $\tau_2$ is the next request epoch after $\tau_1$ $(b)$ only {\it wait} action has been taken in $(\tau_1, \tau_2)$. Suppose, there are $K\geq0$ epochs in $(\tau_1, \tau_2)$. Then, the cost under policy $\pi\in[\tau_1, \tau_2]$ is
}
\[
C_{\pi}([\tau_1, \tau_2]) =c_w (q_1+1)(\tau_2-\tau_1) +C_h(\tau_2-\tau_1)+ c_a \lambda (q_1+2) \tau_2.
\]
Note that, under the policy $\bar{\pi}$,  {\it serve the cached version} action is taken at $\tau_1$, and hence the queue length is zero in $(\tau_1, \tau_2]$. Hence, 
\begin{align*}
  C_{\bar{\pi}}([\tau_1, \tau_2]) &= c_a \lambda (q_1+1) \tau_1 +C_h(\tau_2-\tau_1)+ c_a \lambda \tau_2\\
  &\leq c_a \lambda (q_1+2) \tau_2+C_h(\tau_2-\tau_1)\\ 
  &< C_{\pi}([\tau_1, \tau_2]
\end{align*}
\remove{
\[
C_{\bar{\pi}}([\tau_1, \tau_2]) < C_{\pi}([\tau_1, \tau_2])
\]
}
So, in any renewal cycle, the average cost  under $\bar{\pi}$ is at most the average cost under $\pi$. 
\end{IEEEproof}
Following the above observation, the {\it wait} actions can always be followed by evicting the content as keeping the content in the cache can incur a holding cost  $C_h$ per unit time. \remove{Hence,  if the optimal action is  {\it wait} for additional requests then it is optimal not the {\it serve the cached version} before fetching. Hence, it will be optimal to evict the content while taking the {\it wait} action beacuse keeping the content in the cache incur a waiting cost.

{\color{red}

For $C_h > 0$, it
can be argued that if the optimal action is not to serve the cached version and to wait for additional requests before fetching then it will be optimal to evict the content.} 
{\color{blue}Similar to Proposition~\ref{prop:policy}, we can show that, for $C_h=0$, if the optimal action is not to serve the cached version and to wait for additional requests before fetching, keeping or evicting the cached version incur the same cost.}
}So, we specify Action~$2$ as {\it wait and evict} instead of just {\it wait} as in Figure~\ref{fig:finite_capacity_with_queue}. Moreover, we further segment actions $0$ and $1$ specifying whether content $r$ is kept or evicted after serving. Accordingly, we introduce actions $3$ and $4$ whose annotations are as in Figure~\ref{fig:with_holding_cost_with_queue}. For $r \notin \mathcal{E}_k$ and for $n \in \mathcal{E}_k \setminus \{r\}$, the associated action sets remain as in Figure~\ref{fig:finite_capacity_with_queue}.
We now focus on the optimal caching and delivery problem associated with a tagged content, say content $r$. However, unlike Section~\ref{sec:inf_cap_queue}, content $r$ can be acted  upon even at the request epochs of other contents since keeping the content incurs running cost $C_h$. Hence, as in Section~\ref{sec:sys_model_queue}, we consider all the request epochs to be the decision epochs. We omit the
content indices in the variables as in Section~\ref{sec:inf_cap_queue}. Recall that if $r \in {\cal E}_k$ then its request queue remains empty until {\it wait and evict} action is taken at which point $r$ is evicted. Hence the states
$(Q,\tau,1,1)$ and $(Q,\tau,1,0)$ are not feasible for $Q>0$. So, the states can only take values in $\mathcal{S} \coloneqq \{(0,\tau,1,1),(0,\tau,0,0),(Q,0,1),(Q,0,0),\tau\in\mathbb{R}_+, Q\in\mathbb{Z}_+\}$.
 We can frame the optimal content caching and delivery problem as an MDP with states $s_k \in {\cal S}$ and actions $a_k \in \{0,1,2,3,4\}$ and random noise $(\Delta\tauk,b_{k+1})$ where $ \Delta \tau_k$ and $b_{k+1}$ are Exponential($\beta$) and Bernoulli($p$) random variables, respectively. 
If $s_k=(0,\tau,1,1)$, the next state will be
\begin{align*}
    s_{k+1}=\begin{cases}
        (0, \tau+\Delta\tauk, 1,b_{k+1}) &\text{ if } a_k=0,  \\
        (0,\Delta \tauk,1,b_{k+1})&\text{ if }   a_k=1, \\
        (1,0,b_{k+1}) &\text{ if }a_k=2, \\
        (0,0,b_{k+1})&\text{ if } a_k\in \{3,4\}.
    \end{cases} 
\end{align*}
If $s_k=(0,\tau,1,0)$, the next state will be
\begin{align*}
    s_{k+1}=\begin{cases}
        (0, \tau+\Delta\tauk, 1,b_{k+1}) &\text{ if } a_k=0,  \\
        (0,0,b_{k+1}) &\text{ if }   a_k=2.
    \end{cases} 
\end{align*}
If $s_k=(Q,0,1)$, the next state will be
\begin{align*}
    s_{k+1}=\begin{cases}
        (0,\Delta \tauk,1,b_{k+1}) &\text{ if } a_k=1, \\
        (Q+1,0,b_{k+1}) &\text{ if }a_k=2, \\
        (0,0,b_{k+1})&\text{ if } a_k=3.
    \end{cases} 
\end{align*}
 \begin{table*}[htbp]
\centering
\caption{States and the optimal actions }
\setlength\tabcolsep{2pt}
\label{table:optimal_policy_w_hldg_cost}
\begin{tabular}{|c|c|c|c|c|}
\hline
 Holding   &  \multicolumn{3}{|c|}{Optimal policy ($\pi^\ast$)} & Optimal \\ 
     \cline{2-4}
    {cost ($C_h$)} & $s=\left(0,\tau,1,1\right)$ & $s=\left(0,\tau,1,0\right)$ & $s=\left( Q,0,1\right)$&cost $(\theta)$ \\
     \hline
      $C_h=0 $ & $\pi^{\ast}(s)=\begin{cases}
          0\text{ for } \tau \leq \taustar \\
          2\text{ for } \tau > \taustar,\, Q^{\ast}>0\\
          1\text{ for }\tau>\taustar,\,Q^{\ast}=0
      \end{cases}$ & $\pi^{\ast}(s)=0$ & $\pi^{\ast}(s)=\begin{cases}
          2 \text{ for }Q< Q^{\ast}\\
          1 \text{ for }Q\geq Q^{\ast}
      \end{cases}$ & $\theta=p\beta c_a\lambda\taustar$\\
     \hline
      $ 0<C_h\leq I $&$\pi^{\ast}(s)=\begin{cases}
          0\text{ for } \tau \leq \taub \\
          4\text{ for } \taub<\tau \leq \tautil \\
          2\text{ for } \tau>\tautil,\, \bar{Q}>0\\
          1 \text{ for } \tau>\tautil,\, \bar{Q}=0
      \end{cases}$ & $\pi^{\ast}(s)=\begin{cases}
          0  \text{ for }\tau\leq\taub \\ 
          2 \text{ for }\tau\geq \taub
      \end{cases}$ & $\pi^{\ast}(s)=\begin{cases}
          2 \text{ for }Q< \bar{Q}\\
          1 \text{ for }Q\geq \bar{Q}
      \end{cases}$ & $\theta =p\beta c_a \lambda \tautil$\\
     \hline
    $C_h> I$ & Not applicable & Not applicable
       & $\pi^{\ast}(s)=\begin{cases}
          2 \text{ for }Q< \hat{Q}\\
          3 \text{ for }Q\geq \hat{Q}
      \end{cases}$ & $\theta=\frac{2p\beta c_f+c_w\hat{Q}(\hat{Q}+1)}{2(\hat{Q}+1)}$\\
     \hline
\end{tabular}
\end{table*}
  
Observe from Figure~\ref{fig:with_holding_cost_with_queue} that actions $0$ and $4$ are not applicable when the state is $(Q,0,1)$ and no action is taken when the state is $(Q,0,0)$. 
For a state action pair $(s, a)$, the expected single stage cost $c(s,a)$ will be as follows.
\begin{align}
\lefteqn{c(s,a)=} \nonumber \\
 & \begin{cases}c_a\lambda\tau\mathbbm{1}_{\{a_k\in\{0,4\}\}}+c_f\mathbbm{1}_{\{a_k\in\{1,3\}\}}\nonumber\\
 +\frac{c_w}{\beta}\mathbbm{1}_{\{a_k=2\}}+\frac{C_h}{\beta}\mathbbm{1}_{\{a_k\in\{0,1\}\}} \text{ if }s_k=(0,\tau,1,1)\nonumber\\
 \frac{C_h}{\beta}\mathbbm{1}_{\{a_k=0\}} \text{ if }s_k=(0,\tau,1,0)\nonumber\\
 c_f\mathbbm{1}_{\{a_k\in\{1,3\}\}}+\frac{c_w}{\beta}(Q+1)\mathbbm{1}_{\{a_k=2\}}\nonumber\\
 +\frac{C_h}{\beta}\mathbbm{1}_{\{a_k=1\}} \text{ if }s_k=(Q,0,1)\nonumber\\
 \frac{Qc_w}{\beta} \text{ if }s_k=(Q,0,0).
 \end{cases}   
\end{align}
We can express the MDP problem as 
\begin{equation}
\min_{\pi \in \Pi}J_{\pi}(s) \label{eq:min_J_pi_s}
\end{equation}
where
\begin{equation}
  J_{\pi}(s) = \lim_{T \to \infty}\frac{1}{T}\mathbb{E}\left[\sum_{k=1}^{A(T)}c(s_k,\pi(s_k))\vert s_0=s\right]\label{eq:J_pi_s_with_queue}
\end{equation}
and $\Pi$ is the set of all stationary policies.
The underlying embedded discrete time Markov chain 
is weakly accessible,  and hence, $J_{\pi}(s)$ are independent of $s$
~\cite[Section~4.2]{bertsekas2011dynamic}. Again the Bellman's equations for the continuous time MDP are similar to the discrete-time problems. 
Let $h(s)$ and $\theta$ be the relative cost function and the optimal cost, respectively. Let us define 
\begin{align}
\lefteqn{L_r(\tau){\coloneqq}} \\
& \int_{r}^{\infty}\beta e^{-\beta t}\left(ph(0,t+\tau,1,1)+(1-p)h(0,t+\tau,1,0)\right)dt.\label{eq:lotau_queue}
\end{align}
Then the Bellman's equations for different states are as follows~\cite[Section 5.3]{bertsekas2011dynamic}.
\begingroup
\allowdisplaybreaks
\begin{align}
 h(0,\tau,1,1) = &\min \left\{
     c_a\lambda \tau +\frac{C_h-\theta}{\beta}+\Lotau, \right. \nonumber \\ & c_f+\frac{C_h-\theta}{\beta} +\Lot,\nonumber \\
      &\left. \frac{c_w-\theta}{\beta}+ph(1,0,1)+(1-p)h(1,0,0),\right.\nonumber\\
       & \left. c_f-\frac{\theta}{\beta}+ph(0,0,1)+(1-p)h(0,0,0),\right. \nonumber\\
      & \left. c_a\lambda\tau-\frac{\theta}{\beta}+p h(0,0,1)+(1-p)h(0,0,0)  \right\} \nonumber\\
h(0,\tau,1,0) = &\min \left\{
     \frac{C_h-\theta}{\beta}+\Lotau,\right.\nonumber\\
     &\left. -\frac{\theta}{\beta}+ph(0,0,1)+(1-p)h(0,0,0)\right\},\nonumber\\
 h(Q,0,1) = &\min \left\{
c_f+\frac{C_h-\theta}{\beta}+\Lot, \frac{(Q+1)c_w-\theta}{\beta} \right. \nonumber \\
& +p h(Q+1,0,1)  +(1-p)h(Q+1,0,0),\nonumber\\
&\left.c_f-\frac{\theta}{\beta}+p h(0,0,1)  +(1-p)h(0,0,0)\right\}, \nonumber\\
h(Q,0,0) =& \frac{Qc_w{-}\theta}{\beta}{+}p h(Q,0,1){+}(1{-}p)h(Q,0,0). \label{eqn:bellman_hq00}
  \end{align}
\endgroup

  The following theorem gives the optimal policy for Problem~\eqref{eq:min_J_pi_s}.
\begin{theorem}\label{thm:fin_cap_wait}
     Let us define $\tau^0\coloneqq\frac{2p\beta c_f+c_w\hat{Q}(\hat{Q}+1)}{2p\beta c_a\lambda(\hat{Q}+1)}$ and $I\coloneqq p\beta c_a\lambda\tau^0-pc_a\lambda(1-e^{-\beta\tau^0})$. Let $\taub,\tautil$ and $\bar{Q}$ be the solutions to the following equations.
     \begingroup
     \allowdisplaybreaks
      \begin{align}
    & \beta pc_a \lambda (\tautil \taub{-}\frac{\taub^2}{2}){-}C_h \taub{+}(\bar{Q}+1)c_a \lambda \tautil {-}c_f{-}\frac{c_w\bar{Q}(\bar{Q}+1)}{2p\beta}=0 \label{eq:1_for_uniqueness_wait}\\
     &\beta (\tautil-\taub)+e^{-\beta (\tautil-\taub)}-1-\frac{C_h}{ p c_a \lambda}=0 \label{eq:2_for_uniqueness_wait}\\
     &\bar{Q}=\left\lfloor \frac{p\beta c_a\tautil}{c_w}\right\rfloor \label{eq:3_for_uniqueness_wait}
\end{align}
\endgroup
Further, let $\taustar$ and $ Q^{\ast}$ be the solutions to following equations.
      \begin{align*}
&\tau{=-}\frac{(Q+1)}{p\beta}{+}\frac{1}{p\beta}\sqrt{(Q+1)^2{+}\frac{2p\beta c_f}{c_a\lambda}{+}\frac{Q(Q{+}1)c_w}{c_a\lambda}}, \\
           &Q=\left\lfloor \frac{p\beta c_a\lambda\tau}{c_w}\right\rfloor.    \end{align*}
Finally, let $\hat{ Q}$ be the solution to the following equation. 
\[\hat{ Q}=\left\lfloor\frac{2p\beta c_f+ c_w\hat{Q}(\hat{Q}+1)}{2c_w(\hat{Q}+1)}\right\rfloor.\]
  The optimal policy $\pi^{\ast}$ is as described  in Table~\ref{table:optimal_policy_w_hldg_cost}. 
  \end{theorem}
  \begin{IEEEproof}
    Please see Appendix~\ref{apndx:prf_main_thm}.
\end{IEEEproof}   
The following lemma further characterizes $\taustar,\taub,\tautil,\bar{Q},Q^{\ast},\hat{Q}$ and $\tau^{0}$. We use these characterizations in designing the Whittle index based policy in Section~\ref{eq:indexability and whittle index with waiting}. 
\begin{lemma} 
 \label{uniqueness_lemma_wait}
     $1)$ There exists a unique solution to  \eqref{eq:1_for_uniqueness_wait},~\eqref{eq:2_for_uniqueness_wait}, and~\eqref{eq:3_for_uniqueness_wait} for $\tautil,\taub\in\mathbb{R}_+$ and $\bar{Q}\in\mathbb{Z}_+$.\\ 
  $2)$ $\taub$ and $\tautil$ are decreasing and increasing functions, respectively, of $C_h$ for $0< C_h\leq I$. $\bar{Q}$ is a nondecreasing function of $C_h$ for $0< C_h\leq I$.  Furthermore, $\taub\leq\taustar\leq\tautil\leq\tau^0$ and $Q^{\ast}\leq\bar{Q}\leq \hat{Q}$. 
\end{lemma}
\begin{IEEEproof}
    See Appendix~\ref{apndx:prf_uniqueness_lemma}. 
\end{IEEEproof}

\paragraph*{States~$(Q,\tau,1,b)$ for $Q > 0$} Recall that the optimal policy $\pi^{\ast}$ never leads to the states $(Q,\tau,1,b)$ for $Q > 0$. However, we can extend the domain of $\pi^{\ast}$ to include these states as well. 
Let $\bar{\cal S} \coloneqq {\cal S} \cup \{(Q,\tau,1,b): Q > 0, b\in \{0,1\}\}$.
Let us define 
\begin{equation}
\pi^{\ast}(Q,\tau,1,b) = \begin{cases} \pi^{\ast}(Q,0,1) \ \forall Q > 0 \text{ if } b = 1, \\
2 \ \forall Q > 0 \text{ if } b = 0.
\end{cases}\label{eq:extended_policy}
\end{equation}
We can compute the associated relative costs $h(Q,\tau,1,b)$ and can show that the extended policy $\pi^{\ast}$ and the relative cost function $h$ are solutions to the Bellman's equations for all the states in $\bar{\cal S}$. In particular, the extended  policy $\pi^{\ast}$ yields the optimal actions for all the states in $\bar{\cal S}$. We use this observation while defining Whittle indices in Subsection~\ref{eq:indexability and whittle index with waiting}.

Different contents have different request probabilities~($p_n$) and updated rates~($\lambda_n$). So, the optimal policies for the contents, as given by Theorem~\ref{thm:fin_cap_wait}, have same structure but different parameters, e.g., $Q^{\ast}, \bar{Q}, \tau^{\ast}, \tilde{\tau}$ etc. In the following, we distinguish the parameters and the optimal policies associated with different contents by using superscripts that represent the  contents' indices.     
 \subsection{Whittle index based content caching and eviction}\label{eq:indexability and whittle index with waiting}
  In this section, we introduce the concepts of passive sets and indexability and demonstrate that the contents are indexable. Subsequently, we introduce Whittle indices and present a Whittle index based policy. 

\remove{ Unlike the single content problem with holding cost, in the joint problem a content $r$ can be in state $(Q^r_k,\tauk^r,1,0)$ for some $0 < Q^r_k \leq Q^{\ast r}$. Recall that

the content enters these states due to {\it wait} action at a past decision epoch. In this case  

Hence $W^r(Q^r_k,\tauk^r,1,0) = 0$ for all $Q^r_k > 0$.
We argue that  

Observe that }

\paragraph*{Passive sets}
The passive set for a content $n$ for a given holding cost $C_h$, $\mathcal{P}^n(C_h)$, is the set of states in which the optimal action does not require the content to be cached. More precisely, 
\[\mathcal{P}^n(C_h):=\left\{s\in\bar{{\cal S}}: \pi^{\ast n}(s)\in\{2,3,4\}\right\}\] 
where $\bar{{\cal S}}$ is the extended state space for single content problem as defined in Section~\ref{sec:single-content-holding-cost}.

\paragraph*{Indexability}
 A content $n$ is  called indexable if its passive set  is nondecreasing in $C_h$~\cite{ayesta2021computation}. More specifically,
 \[\mathcal{P}^n(C_{h_1}) \subseteq \mathcal{P}^n(C_{h_2}) \text{ for } C_{h_1} \leq C_{h_2}. \]
  The RMAB problem under consideration is called indexable if every content is indexable.
  
\paragraph*{Whittle index} 
The whittle index associated with state $s$ of content $n$, $W^n(s)$, is the minimum holding cost for which $s$ is in the passive set. In other words, 
\[W^n(s)=\min\{C_h:s \in \mathcal{P}^n(C_h)\}.\]   
\remove{Since the Whittle indices $W^n(s)$ are determined for the contents in states $(0,\tau,1,0)$ and $(q,0,1)$ we establish indexability for the restricted state space $\mathcal{S}_2=\{(0,\tau,1,0),(q,0,1),\tau\in\mathbb{R}_+, q\in\mathbb{Z}_+\}$.} 
\begin{figure}
\begin{subfigure}[t]{0.5\linewidth}
\centering  
\includegraphics[width=\linewidth]{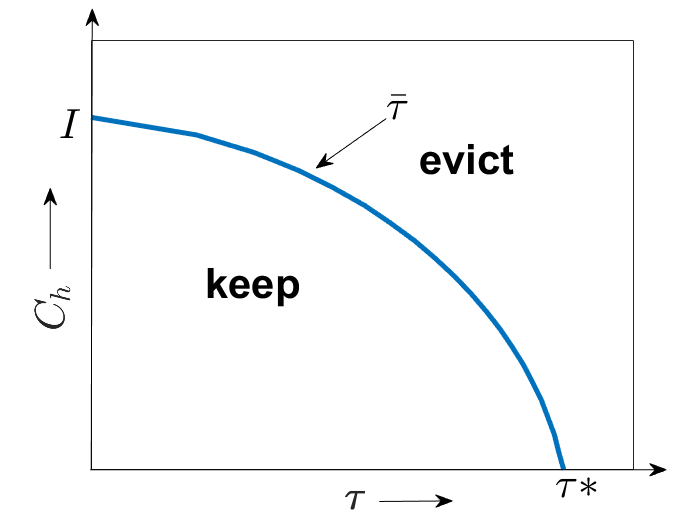}
   \caption{$s=(0,\tau,1,0)$ }
   \label{fig:h_0_tau_1_0_}
\end{subfigure}
\begin{subfigure}[t]{0.49\linewidth}
\centering  
\includegraphics[width=\linewidth]{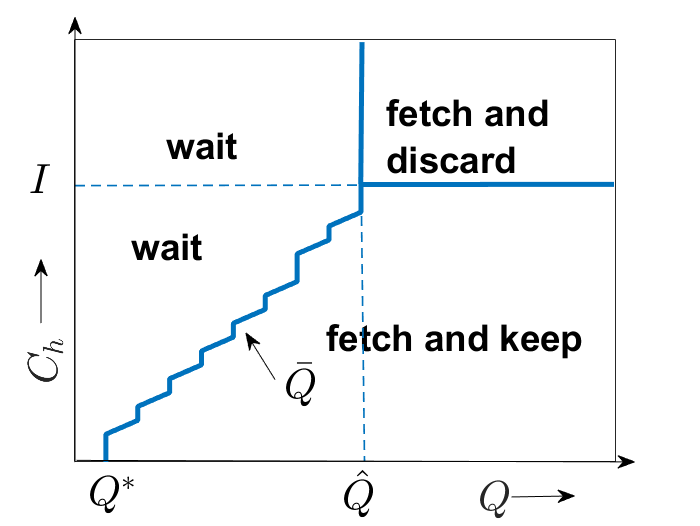} 
   \caption{$s=(Q,0,1)$}
   \label{fig:h_q_0_1}
\end{subfigure}
\caption{Optimal policy structure with respect to $C_h$ for states $(0,\tau,1,0)$ and $(Q,0,1)$; $\taustar, Q^{\ast}, \taub, \bar{Q}, \hat{Q}$, and $I$ are as in Theorem~\ref{thm:fin_cap_wait}.}
\label{fig:plot_for_optimal_action}
\end{figure}
We first prove that all the contents are indexable.

\begin{lemma}
Each content $n$ is indexable.
\end{lemma}
\begin{IEEEproof}
\remove{We write passive sets for content $n$ for different holding costs $C_h$ using Theorem~\ref{thm:fin_cap_wait} and~\eqref{eq:extended_policy}.
From~\eqref{eq:extended_policy},
$\pi^{\ast}(s)=2$ for all $s\in\{(Q,\tau,1,0): Q>0,  \tau\geq 0\}$ for all $C_h\geq 0$. Furthermore, for $C_h=0$, from Theorem~\ref{thm:fin_cap_wait} and~\eqref{eq:extended_policy},
$\pi^{\ast}(s)=2$ for all $s\in\{(Q,\tau,1,1), (Q,0,1): Q<Q^{\ast}\}$. Hence, 
\begin{align*}
    {\cal P}^n(0)= &\left\{(Q,\tau,1,1),(Q,0,1): Q<Q^{\ast} \right\}\\
    &\cup\left\{(Q,\tau,1,0),(Q,0,0):Q>0,\tau\geq 0 \right\}.
\end{align*}
Following similar arguments, for $0<C_h<I$, 
\begin{align*}
\mathcal{P}^n(C_h)=&\left\{(0,\tau,1,0), (Q,0,1): \tau \geq \taub(C_h), Q\leq \bar{Q}(C_h)\right\} \\
& \cup\left\{(Q,\tau,1,0) ,(Q,0,0):Q>0,\tau\geq 0\right\}\\
&\cup\left\{(Q,\tau,1,1):Q\leq \bar{Q}(C_h),\tau\geq 0\right\}.
\end{align*} 
 where we write $\taub(C_h)$ and $\bar{Q}(C_h)$ to explicitly show the dependence of $\taub$ and $\bar{Q}$ on $C_h$. From Lemma~\ref{uniqueness_lemma_wait}, $\taub(C_h)$ and $\bar{Q}(C_h)$ are decreasing and nondecreasing functions of $C_h$, respectively. Hence ${\cal P}^n(C_h)$ is nondecreasing set for $0\leq C_h\leq I$. Finally, for $C_h>I$, $\mathcal{P}(C_h)=\bar{\cal S}$, and hence, $\mathcal{P}(\infty)=\bar{\cal S}$. 

Clearly, $\mathcal{P}^n(C_h)$ is nondecreasing for all $C_h \geq 0$.
Hence, content $n$ is indexable.}
Please see Appendix~\ref{proof:indexability_wait}.
\end{IEEEproof}

\paragraph*{Whittle index based policy}
Suppose the requested content at $t_k$ is not cached, i.e., $R_k\notin \mathcal{E}_k$, and the local cache chooses to fetch $R_k$, then the local cache must also choose which $M$ of the $M+1$ contents ${\cal E}_k \cup \{R_k\}$ to keep. The state of each content $n \in \mathcal{E}_k$ is $(Q_k^n,\tau_k^n,1,0)$ whereas that of $R_k$ is $(Q_k^{R_k},0,1)$. Under the Whittle index based caching policy, the Whittle indices for all the contents in ${\cal E}_k \cup \{R_k\}$ for their respective states are computed and the content with the least index is not cached or is evicted if it is readily cached. We thus require the Whittle indices for all the contents for the states of the forms $(Q,\tau,1,0)$ and $(Q,0,1)$. The following theorem provides these indices. 

\remove{
We plotted the optimal actions for the single content problem in Figure~\ref{fig:plot_for_optimal_action} for the states $(Q,\tau,1,0)$ and $(Q,0,1)$. From Figure~\ref{fig:h_q_0_1}, note that the optimal action for the content $R_k$  is $2$ or {\it wait} if $Q^{R_k}\leq Q^{\ast R_k}$. Hence, the local cache need to choose $M$ contents only if the queue length of $R_k$ is more than $Q^{\ast R_k}$. 
Furthermore, if $Q_k^n>0$ for any content $n \in \mathcal{E}_k$, then it means that the local cache will never serve the cached content when the content is requested, the local cache will wait or fetch a fresh version. This implies the contents in $\mathcal{E}_k$ with $Q_k^n>0$ can be evicted and can be replaced by $R_k$. Hence, the local cache only needs to choose $M$ contents if $Q_k^n=0$ for all the contents in $\mathcal{E}_k$ and $Q^{R_K}>Q^{\ast R_k}$.}
\remove{
Subsequently, we propose a policy that discards the content with the lowest Whittle index; thus, the base station chooses the $M$ contents to retain based on the Whittle index policy.  Towards this, we first demonstrate that each content in $\mathcal{E}_k\cup R_k$ is indexable and compute Whittle indices for each content in $\mathcal{E}_k\cup R_k$.

The following Theorem provides the Whittle indices for different contents. }
\begin{theorem}
\label{thm:whittle_indices_with_queue}
$(a)$ For $n\in\mathcal{E}_k$, 
    \[W^n(Q,\tau,1,0)=\begin{cases}
        C_h(\tau) \text{ for }\tau\leq {\taustar}^n \text{ and }Q=0,\\
        0 \text{ for }\tau> {\taustar}^n \text{ or }Q>0
    \end{cases}\]
    where ${\taustar}^n$ is as defined in Theorem~\ref{thm:fin_cap_wait} and $C_h(\tau)$, $\tautil$, and $\bar{Q}$ are the solutions to the following equations~(\eqref{eq:1_for_whittle_index},~\eqref{eq:2_for_whittle_index} and~\eqref{eq:3_for_whittle_index} are three equations in three unknowns  $C_h(\tau)$, $\tautil$, and $\bar{Q}$ though $\tautil$, and $\bar{Q}$ are not needed for $W^n(Q,\tau,1,0)$).
    \begin{align}
     & \beta p_nc_a \lambda_n (\tautil \tau-\frac{\tau^2}{2})-C_h(\tau) \tau{+}(\bar{Q}+1)c_a \lambda_n \tautil \nonumber\\
     &\hspace{2.2 cm}-c_f-\frac{c_w\bar{Q}(\bar{Q}+1)}{2p_n\beta}=0 \label{eq:1_for_whittle_index}\\
      &\beta (\tautil-\tau)+e^{-\beta (\tautil-\tau)}-1-\frac{C_h(\tau)}{ p_n c_a \lambda_n}=0 \label{eq:2_for_whittle_index}\\
     &\bar{Q}=\left\lfloor \frac{p_n\beta c_a\tautil}{c_w}\right\rfloor.\label{eq:3_for_whittle_index}
    \end{align}
$(b)$ For $n = R^k$,  
 \[W^n(Q,0,1)=\begin{cases}
        0 \text{ for }Q< Q^{\ast n},\\
        C_h(Q) \text{ for } Q^{\ast n} \leq Q< \hat{Q}^{n},\\
        I^{n} \text{ for }Q\geq \hat{Q}^{n},
    \end{cases}
    \]
    where $Q^{\ast n},\hat{Q}^n$ and $I_n$ are as defined in Theorem~\ref{thm:fin_cap_wait}, and $C_h(Q)$, $\taub$ and $\tautil$ are solutions to~\eqref{eq:1_for_whittle_index},~\eqref{eq:2_for_whittle_index} and~\eqref{eq:3_for_whittle_index} with $C_h(\tau),\bar{Q}$ and $\tau$  are replaced with $C_h(Q),Q$ and $\taub$, respectively~($\tautil$ and $\bar{Q}$ are not needed for $W^n(Q,\tau,1,0)$). 
\end{theorem}
\begin{IEEEproof}
  Recall that $W^n(Q,\tau,1,0)=\min\{C_h: \pi^{\ast n}(Q,\tau,1,0)=2\}$ and $W^n(Q,0,1)=\min\{C_h:\pi^{\ast n}(Q,0,1)\in\{2,3\}\}$. The expressions follow from Theorem~\ref{thm:fin_cap_wait}.
\end{IEEEproof}

\subsection{Content caching and delivery policy}
\label{sec:content-caching-delivery-policy}
We now describe the Whittle index based content fetching, caching and delivery policy. Towards this, let us revisit the set of available actions as shown in Figure~\ref{fig:finite_capacity_with_queue}.

If the requested content at $t_k$ is cached, i.e., if $R_k\in \mathcal{E}_k$, we need not evict any content in $\mathcal{E}_k$. 
 \remove{
 In this case, $R_k$ is in state $(Q_k^{R_k},\tauk^{R_k},1,1)$ whereas $n \in \mathcal{E}_k \setminus R_k$ are in states  $(Q_k^n,\tau_k^n,1,0)$. Hence, the optimal decision will be to keep content $R_k$ and the other contents in ${\cal E}_k$.
 }
 The optimal action for $R_k$ in this case is as given by  Theorem~\ref{thm:fin_cap_wait} with $C_h=0$. More specifically, it is as follows.
\begin{enumerate}
    \item If $\tauk^{R_k} \leq \tau^{\ast R_k}$, {\it serve the cached copy},
    \item if $\tauk^{R_k} > \tau^{\ast R_k} $,
    \begin{enumerate}[(a)]
        \item if $Q_k^{R_k} < Q^{\ast R_k}$, {\it do not serve}~({\it wait}),
        \item if $Q_k^{R_k} \geq Q^{\ast R_k}$, {\it fetch, serve and cache}.
    \end{enumerate}
\end{enumerate}
where ${\taustar}^{R_k}$ and $Q^{\ast R_k}$ are as defined in Theorem~\ref{thm:fin_cap_wait}.

If the requested content is not in the cache, i.e., $R_k\notin \mathcal{E}_k$, the content caching and eviction decisions are taken based on the concerned contents' Whittle indices as described in Section~\ref{eq:indexability and whittle index with waiting}. 
Recall that the state of each content $n$ in $\mathcal{E}_k$ is $(Q_k^n,\tau_k^n,1,0)$ and that of  $R_k$ is $(Q_k^{R_k},0,1)$.
The optimal actions for the contents ${\cal E}_k \cup \{R_k\}$ are as follows.
\begin{enumerate}
    \item If $Q^{R_k}< Q^{\ast R_k}$, {\it do not serve} $R_k$ and {\it keep} all $n \in {\cal E}_k$,
    \item If $Q^{R_k}\geq Q^{\ast R_k}$, compute $W^{R_k}(Q_k^{R_k},0,1)$ and $W^n(Q_k^n,\tau_k^n,1,0)$ for all $n \in {\cal E}_k$ using Theorem~\ref{thm:whittle_indices_with_queue}, 
    \begin{enumerate}[(a)]
    \item if $W^{R_k}(Q_k^{R_k},0,1) > \min_{n \in {\cal E}_k} W^n(Q_k^n,\tau_k^n,1,0)$, {\it fetch, serve and cache} $R_k$ and evict a $n \in \arg\min_{n \in {\cal E}_k} W^n(Q_k^n,\tau_k^n,1,0)$,
    \item if $W^{R_k}(Q_k^{R_k},0,1) \leq \min_{n \in {\cal E}_k} W^n(Q_k^n,\tau_k^n,1,0)$, {\it keep} all $n \in {\cal E}_k$ and
    \begin{enumerate}[(i)]
        \item if  $Q^{R_k}< \hat{Q}^{R_k}$,  {\it do not serve} $R_k$,
        \item if $Q^{R_k}\geq \hat{Q}^{R_k}$,  {\it fetch, serve and discard} $R_k$.
    \end{enumerate}
    \end{enumerate}
\end{enumerate}

\remove{
\begin{algorithm}
\caption{Whittle Index-based Caching}
\label{alg:whittle_index_policy_with_queue}
\begin{algorithmic}
\State Initialization: Start with an arbitrary ${\cal E}_0$, $\{\tau_0^{n}:n\in {\cal E}_0\}$,and $\{Q_0^n:n\in[N]\}$
\Procedure{Caching}{ }
    \ForAll{$t_k$}
              \If {$R_k \in \mathcal{E}_k$} 
                      \If{$\tauk^{R_k}\leq \tau^{\ast R_k}$ } 
                      \State  serve the cached copy and keep,
                      \ElsIf{$Q_k^{R_k}<=Q_k^{\ast R_k}$}
                      \State wait
                       \Else
                             \State   fetch and cache the requested content,  
                      \EndIf

             \Else
                 \ForAll{$n \in {\cal E}_k$}
                         \State calculate $W^n(Q^n_k,\tau^n_k,1,0)$  as in Theorem~\ref{thm:whittle_indices_with_queue}
                      \EndFor
                      \State calculate $W^{R_k}(Q_k^{R_k},0,1)$  as in Theorem~\ref{thm:whittle_indices_with_queue}
                      \If{ $W^{R_k}(Q_k^{R_k},0,1)$ $ > \min_{n \in {\cal E}_k}W^n(Q^n_k,\tau^n_k,1,0)$}
                      \State fetch and cache $R_k$ 
                      \ElsIf{$Q_k^{R_k}<\hat{Q}^{R_k}$} 
                      \State wait
                      \Else 
                      \State fetch $R_k$, serve and do not cache      
                      \EndIf
             \EndIf
   \EndFor
\EndProcedure
\end{algorithmic}
\end{algorithm}
}

\begin{remark}[Asymptotic optimality] 
As mentioned in Section~\ref{sec:finite_cache_capacity_with_waiting}, the proposed Whittle index based policy becomes asymptotically optimal  as the number of contents $N$ approaches
infinity and the cache capacity $M$ also grows proportionately.
\end{remark}

\begin{remark}[Impatient requests]
For sufficiently large $c_w$, the optimal policy for the single content problem with holding cost (Theorem~\ref{thm:fin_cap_wait}) becomes same as~\cite[Theorem 2]{10723415} where one does not have the option ({\it not to serve the requested contents}~({\it wait}). Consequently, our proposed Whittle index based content caching and delivery policy becomes same as  in~\cite{10723415}[Algorithm~1]. In general allowing {\it wait} action for the requested contents results in smaller average costs~(see Figure~\ref{fig: vary_c_w_queue}.
\end{remark}
\section{Numerical Results}
Now we numerically evaluate the performance of the Whittle index-based content caching and delivery policy derived in Section~\ref{sec:content-caching-delivery-policy}. We consider $N=1000$ content with the aggregate request arrival rate $\beta=40$ and {\em Zipf's} content popularity distribution with parameter $1$. In other words, we assume $p_n = \frac{1/n}{\sum_{i=1}^N 1/i}$ for all $n$. We assume identical content update rate $\lambda_n=0.01$ for all the contents at the server. Moreover, we assume the ageing cost per unit version age $c_a=0.1$, the fetching cost $c_f=1$ and the waiting cost per unit time $c_w=0.01$. 

\subsection{Optimality}\label{num:optimality}
We compare the average cost of the proposed Whittle index based policy and the optimal cost of the relaxed RMAB problem (\eqref{eq:POMDP_with_finite_cache} subject to~\eqref{relaxed_constraint_fnite_cache}). The latter cost is  a lower bound on the optimal cost of the content caching and delivery problem and can be numerically computed  using Theorem~\ref{thm:fin_cap_wait}. The gap between the cost of the proposed policy and the optimal cost of the relaxed RMAB problem is an upper bound on the suboptimality gap for the former. 

We vary the cache capacity $M$ from $200$ to $300$ and demonstrate for each $M$ that the cost of the proposed policy is indeed very close to the optimal cost of relaxed RMAB problem~(see Figure~\ref{fig: comparison_with_relaxed_myopic}). We can thus infer that  the proposed  Whittle index based policy performs very close to the optimal policy. 
\subsection{Comparison to a Myopic Policy}
\begin{figure}[h]
\begin{subfigure}[t]
{0.49\linewidth}
  \centering
\includegraphics[width=1\linewidth]{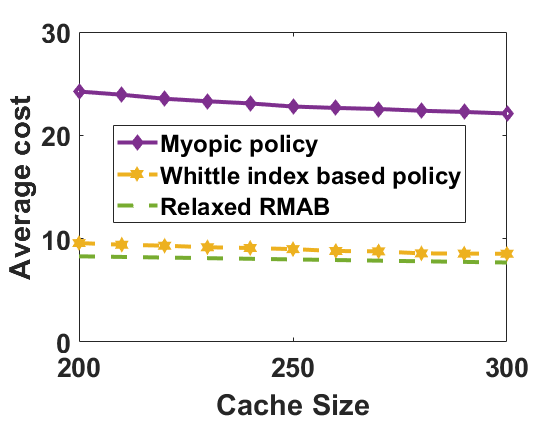}
\caption{}
%\caption{Average cost vs cache size for different values of $c_W$. } 
\label{fig: comparison_with_relaxed_myopic}
\end{subfigure}
\begin{subfigure}[t]
{0.5\linewidth}
  \centering
\includegraphics[width=1\linewidth]{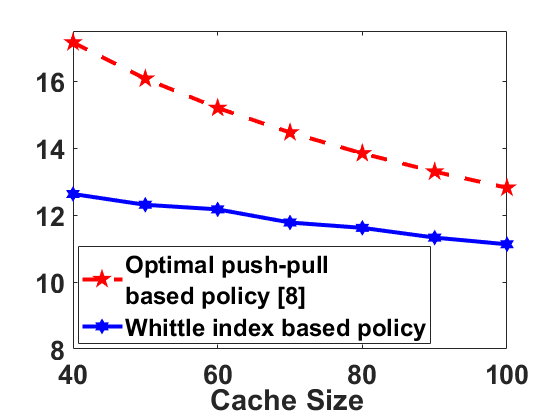}
%\caption{Average wait time vs $c_w$ for cache size $100$}
\caption{}
\label{fig: comparison_with_push_pull}
\end{subfigure}
\caption{Performance of the Whittle index based policy. For each cache size, its cost is close to the optimal value of the relaxed RMAB problem and is much smaller than the cost of a Myopic and push-pull based caching policy~\cite{10452408}.
}
\label{fig: comparison_queue}
 \end{figure}

In Figure~\ref{fig: comparison_with_relaxed_myopic}, we also compare the average cost of the proposed Whittle index based policy to that of a {\it Myopic policy} based on~\cite[Section 4.3] {bertsekas2012dynamic}, which at each decision epoch takes an action to minimize the single stage and terminal costs, considering the next state as the terminal state. Hence, the requested contents in the current and next slots must be served before the next slot ends. 

Let $R_k$ be the content requested at $t_k$. One of the following two cases can happen.   
\paragraph{$R_k\in{\cal E}_K$} 
Then the myopic policy prescribes the action $a^\ast_k\in \arg\min_jC_{j,k}$ where $C_{j,k}$s are the costs corresponding to the actions $j \in \{0,1,2\}$~(see Figure~\ref{fig:finite_capacity_with_queue}).  Let us define, \[w_k\coloneqq\sum_{l\in \mathcal{E}_k\setminus R_k}\frac{Q^lc_w}{\beta}+\sum_{l\in \mathcal{E}_k\setminus R_k}p_l\min\{c_f+\frac{(Q^l+1)c_w}{\beta}.\]It can be seen that
\begingroup\allowdisplaybreaks \begin{align*}
    C_{0,k} = & c_a\lambda\tau^{R_k}(Q^{R_k}+1)+p_{R_k}\min\left\{c_f,c_a\lambda(\tau^{R_k}+\frac{1}{\beta})\right\}\\
    &+w_k,\\
       C_{1,k}= &c_f+p_{R_k}\min\left\{c_f,\frac{c_a\lambda}{\beta}\right\}+w_k,\nonumber\\
       C_{2,k}= & \frac{c_w(Q^{R_k}{+}1)}{\beta}{+}p_{R_k}\min \Big\{c_f,(Q^{R_k}{+}2)c_a\lambda(\tau^{R_k}+\frac{1}{\beta})\Big\}\nonumber\\
       &+(1-p_{R_k})\min\left\{c_f{+}(Q^{R_k}{+}1)c_a\lambda(\tau^{R_k}{+}\frac{1}{\beta})\right\}+w_k.
    \end{align*}
    \endgroup
\paragraph{$R_k\notin{\cal E}_k$} 
Then the myopic policy prescribes the action $a^\ast_k\in \arg\min_jC'_{j,k}$ where $C'_{j,k}$s are the costs corresponding to the actions $j \in \{1,2,3\}$~(see Figure~\ref{fig:finite_capacity_with_queue}). It can be seen that
\begingroup \allowdisplaybreaks\begin{align*}
    C'_{1,k}=&c_f{+}\min_{n\in \mathcal{E}_k}\bigg\{p_nc_f{+}\\&\sum_{l\in\mathcal{E}_k\setminus n}p_l\min\Big\{c_f,(Q^l{+}1)c_a\lambda(\tau^l{+}\frac{1}{\beta})\Big\}\bigg\}\nonumber\\
    &+p_{R_k}\min\{c_f,\frac{c_a\lambda}{\beta}\},\\
     C'_{2,k}{=}&\frac{c_w(Q^{R_k}{+}1)}{\beta}\\
     &{+}\sum_{n\in \mathcal{E}_k}p_n\left(\min\Big\{c_f,(Q^n{+}1)c_a\lambda(\tau^n{+}\frac{1}{\beta})\Big\}\right).\\
 C'_{3,k}=&c_f{+}p_{R_k}c_f+{\sum_{n\in \mathcal{E}_k}p_n}\left(\min\Big\{c_f,(Q^n+1)c_a\lambda(\tau^n{+}\frac{1}{\beta})\right).
\end{align*}
\endgroup
If action~$1$ is taken for $R_k$ then one of the contents in ${\cal C}(t_k)$ needs to be evicted. The myopic policy prescribes to evict a content $n'$  where 
\begin{align*}n' \in & {\arg\min}_{n\in \mathcal{E}_k}\bigg\{p_nc_f\\
&+\sum_{l\in\mathcal{E}_k\setminus n}p_l\min\Big\{c_f,(Q^l+1)c_a\lambda(\tau^l{+}\frac{1}{\beta})\Big\}\bigg\}.
\end{align*}

This comparison shows that the proposed policy significantly outperforms the Myopic policy.
\subsection{Comparison to a push-pull based policy~\cite{10452408}}
In Figure~\ref{fig: comparison_with_push_pull}, we also compare the average cost of the proposed Whittle index based policy to that of a push-pull based policy~\cite{10452408}. For this we consider the same parameters as in Section~\ref{num:optimality} and vary the cache size from $40$ to $100$. The policy in~\cite{10452408} is static; it gives a fixed set of contents to be cached and a protocol to update these in the cache. If a requested content is not in the cache, the cache fetches and serves the content and discards it after serving, incurring a fetching cost for each such request. In contrast, our policy is dynamic and also introduces a {\it wait} option. This allows the system to pool requests and to reduce the instances of content fetches, reducing the fetching cost. Our policy significantly reduces the overall cost compared to the static policy~\cite{10452408}.

\subsection{Effect of waiting cost $c_w$}
\begin{figure}[h]
\begin{subfigure}[t]
{0.5\linewidth}
  \centering
\includegraphics[width=\linewidth]{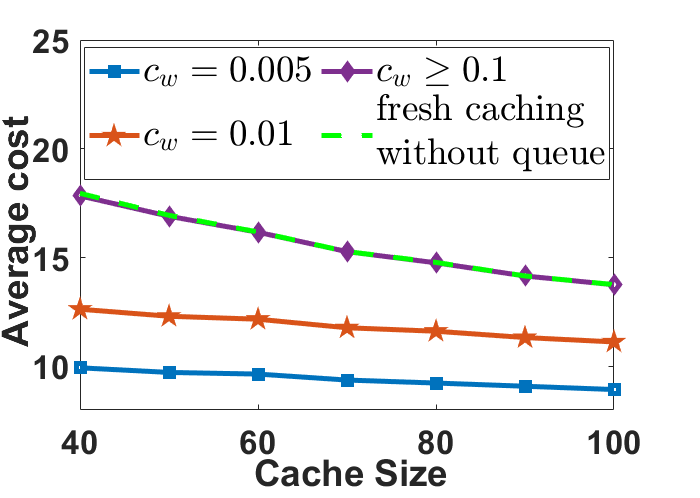}
\caption{}
%\caption{Average cost vs cache size for different values of $c_W$. } 
\label{fig: vary_c_w_queue}
\end{subfigure}
\begin{subfigure}[t]
{0.49\linewidth}
  \centering
\includegraphics[width=\linewidth]{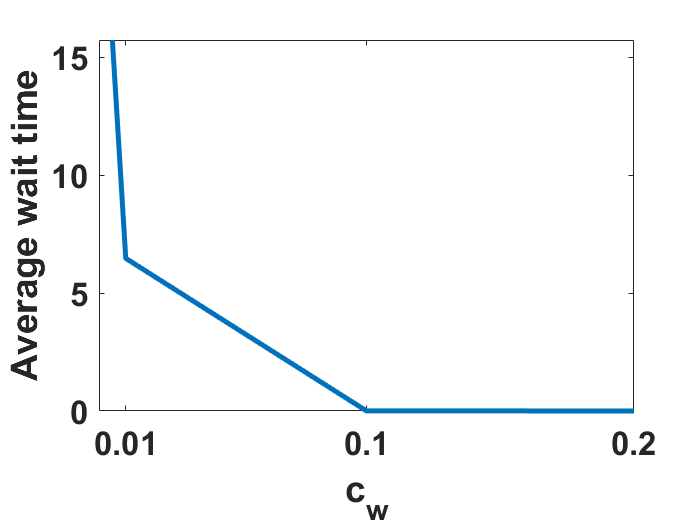}
%\caption{Average wait time vs $c_w$ for cache size $100$}
\caption{}
\label{fig: vary_c_w_changes_queue}
\end{subfigure}
\caption{ Impact of $c_w$ on average cost and waiting time. $(a)$ The average cost increases as $c_w$ increases but it becomes oblivious to $c_w$  beyond $c_w = 0.1$. $(b)$ The average wait time reduces as $c_w$ increases is very close to zero for $c_w \geq 0.1$.
}
 \end{figure}

We vary the waiting cost $c_w$ over $\{0.005,0.01,0.1\}$ and plot the average cost vs cache size for each value of $c_w$ in Figure~\ref{fig: vary_c_w_queue}. We see that  the average cost increases withe $c_w$; the change is noticeable as $c_w$ increases from $0.005$ to $0.1$ but is not noticeable as $c_w$ increases beyond $0.1$. This behavior can be understood as follows.
As $c_w$ increases, the average waiting time  of the requests decreases and as $c_w$ reaches $0.1$, the latter becomes very close $0$~(see Figure~\ref{fig: vary_c_w_changes_queue}). Consequently, 
the average waiting cost is close to zero for $c_w \geq 0.1$. For higher values of $c_w$, the optimal actions are confined to {\it serve the cached version} and {\it fetch and serve}, and hence the optimal cost is not sensitive to the exact value of $c_w$.

In Figure~\ref{fig: vary_c_w_queue}, we also see that the average cost for $c_w\geq0.1$ almost equals the average cost without {\it wait} option~\cite{10723415}.

\section{Conclusion}
We studied a content fetching, caching and delivery problem in a system with a finite capacity local cache. We accounted for the content fetching, ageing and waiting costs and aimed at minimizing the long term average cost.
We first provided the optimal policy under the assumption of infinite cache capacity~(Theorem~\ref{thm:inf_cap_wait}). Subsequently, we formulated the finite cache capacity problem as a RMAB and offered a Whittle index based policy~(Section~\ref{sec:content-caching-delivery-policy}).  We also explicitly characterized the Whittle indices~(Theorem~\ref{thm:whittle_indices_with_queue}).
We numerically demonstrated that the proposed is closed to optimal and substantially outperforms a Myopic and push-pull based policy~\cite{10452408}.

A potential future direction is to consider wireless links between the server and the local cache and between the local cache and the users. We would like to extend this study to a network of local caches each having its own user base.

% \input{Indexability}
% \input{Numerical Results}
%\input{single content with infinite caching capacity with missing cost and wireless link}
%\section{Single content with infinite caching capacity and holding cost}
%\input{Unreliable channel with holding cost}
%\input{indexability_with_wireless}
%\nocite{*}
\bibliographystyle{ieeetr}
\bibliography{reference}
\appendix
\subsection{Proof of Theorem~\ref{thm:inf_cap_wait}}
\label{apndx:prf_inf_cap}

\begin{lemma}
\label{eq:non_decreasing_in_q_tau}
  For any $q\in\mathbb{Z}_+$ and $t\in\mathbb{R}_+$, \begin{equation*}
      h(Q+q,\tau+t)\geq h(Q,\tau)\, \forall Q\in\mathbb{Z}_+,\tau\in\mathbb{R}_+. 
  \end{equation*}
  \end{lemma}
  \begin{IEEEproof}
      The proof of this lemma  uses induction and    relative value iteration and is standard~\cite{bertsekas2011dynamic}. 
  \end{IEEEproof}
  The following lemma characterize the optimal policy. Let $\pi^{\ast}$ denote the optimal policy. 
  \begin{lemma} There exists a $\taustar$ such that
      \[\pi^{\ast}\begin{cases}
          =0, \text{ for }\tau\leq\taustar,\\
          \in\{1,2\}, \text{ otherwise. }
      \end{cases}\]
  \end{lemma}
  \begin{IEEEproof}
       We note that, $h(q,0)=-\frac{\theta}{\beta}+\int_{0}^{\infty}\beta e^{-\beta t}h(0,t)dt$. Hence, for $\tau=0$, it is always optimal to serve the cached copy irrespective of the value of the queue length $(q)$. Note that, once the cached copy is served, the queue length becomes $0$.
      Let us define \begin{align*}
          h&\coloneqq c_f-\frac{\theta}{\beta}+\int_{0}^{\infty}\beta e^{-\beta t}h(0,t)dt\\
          x(q,\tau)&\coloneqq c_a\lambda\tau+\int_{0}^{\infty}\beta e^{-\beta t}h(0,t+\tau)dt\\
          y(q,\tau)&\coloneqq -\frac{\theta}{\beta}+\frac{c_w}{\beta}(q+1)+\int_{0}^{\infty}\beta e^{-\beta t}h(q+1,t+\tau)dt.
      \end{align*} 
      Since $h(q,\tau)$ is non decreasing in $q$ and $\tau$, $x(0,\tau)$ is strictly increasing in $\tau$. Hence, there exists a $\taub>0$ such that for all $\tau\geq \taub$, $x(0,\tau)\geq h$.  Furthermore, there exists a $\tau'>0$ such that for all $\tau\geq\tau'$ $x(0,\tau)>y(0,\tau)$. 
      Let us define $\taustar=\min\{\taub,\tau'\}$.
      Note that if $\taustar\leq\tau'$, then
      \[\pi^{\ast}=\begin{cases}
          0, \text{ for }\tau\leq\taustar,\\
          1, \text{ otherwise. }
      \end{cases}\]
      Let us consider $\tau'>\taustar$.  Consider a policy $\pi$ such that $\pi(q_1,\tau_1)=2$ and $\pi(q_2,\tau_2)=0$ for $\taustar\leq\tau_1<\tau_2<\taub$.  Then the cost under policy $\pi$ in $[\tau_1,\tau_2]$ is \[c_{\pi}([\tau_1 \,\tau_2])=\frac{(q_1+1)c_w}{\beta}+(q_2+2)c_a\lambda \tau_2.\]
Note that, $q_2>q_1$. 
      Let us consider a policy $\bar{\pi}$ such that $\bar{\pi}(q_1,\tau_1)=\bar{\pi}(q',\tau_2)=0$. Note that $q'=0$. Then the cost under policy $\pi$ in $[\tau_1,\tau_2]$ is 
      \begin{align*}
          c_{\bar{\pi}}([\tau_1 \,\tau_2])&=(q_1+1)c_a\lambda\tau_1+c_a\lambda \tau_2\\&\leq+(q_2+2)c_a\lambda \tau_2\\&\leq c_{\pi}([\tau_1 \,\tau_2])
      \end{align*} 
      Hence the optimal policy belongs to a policy class where if the optimal action is {\it wait} for some epoch then the optimal action is to wait until {\it fetch}. 
       \end{IEEEproof}
       Let us consider the times beyond $\taustar$. The following lemma states that for $\tau>\taustar$ the optimal policy is a function of the queue length $Q(\tau)$ only. 
       \begin{lemma}\label{lemma:threshold_policy}
           The optimal policy \[\pi^{\ast}(q,\tau)=\begin{cases}
              0 \text{ for }\tau\leq \taustar \nonumber\\
              2 \text{ for }q< Q^{\ast} \text{ and }\tau\geq\taustar \nonumber\\
              1 \text{ for }q\geq Q^{\ast}.  \nonumber
          \end{cases}\]
          where $\Qstar$ and $\theta$ are the solutions to the following equations. 
          \begin{align}
    \frac{c_w\Qstar(\Qstar+1)}{2\beta}-\frac{(\Qstar+1)\theta}{\beta}+c_f=0.\label{eq:first_eq}\\
     \Qstar=\left\lfloor \frac{\theta}{c_w} \right\rfloor. \label{eq:2nd_eq}
    \end{align}
          \end{lemma}
\begin{IEEEproof}
First we will consider a policy as follows $\pi(q,\tau)=0$ for $\tau\leq\taustar$.   For $\tau>\taustar$ and for some $\Qstar\geq 0$ 
\[\pi(Q, \tau) =
\begin{cases} 
2 & \text{for } Q < \Qstar \\
1 & \text{for } Q \geq \Qstar
\end{cases}
\]
Then we shall compute the relative cost functions and average cost under the policy $\pi$. Then we will show that the relative cost functions under policy $\pi$ satisfy Bellman's equation~\eqref{bellman_eqn_with_queue_infinite_capacity}. 

Let $h_\pi(Q,\tau)$ be the value function at $(Q,\tau)$ and $\theta_\pi$ be the average cost under policy $\pi$. 
From the policy $\pi$, $h_\pi(Q,\tau)$ depends on the values of $Q$ only. 
We define \begin{equation}
    g(Q)=h_\pi(Q,\tau) \label{eq:def_g_Q}
\end{equation} 
Hence according to the policy $\pi$ for $Q<\Qstar$,
\begin{align}
    g(Q)=&-\frac{\theta_\pi}{\beta} + \frac{c_w(Q+1)}{\beta} + \int_{0}^{\infty} \beta e^{-\beta t} g(Q+1)dt\nonumber\\
    =&-\frac{\theta_\pi}{\beta} + \frac{c_w(Q+1)}{\beta} +  g(Q+1).\label{eq:g_q_below_Qstar}
\end{align}
For $Q\geq\Qstar$, 
\begin{align}
g(Q)=&-\frac{\theta_\pi}{\beta} + c_f + \int_{0}^{\infty} \beta e^{-\beta t} g(0)dt\nonumber\\
    =&-\frac{\theta_\pi}{\beta} + c_f+  g(0)\nonumber\\
    \coloneqq& h_\pi. \label{eq:g_q_beyond_Qstar}
\end{align}
From~\eqref{eq:g_q_below_Qstar}, 
\begin{align}
    g(\Qstar-1)=&-\frac{\theta_\pi}{\beta} + \frac{c_w\Qstar}{\beta} +  g(\Qstar)\nonumber\\
    =&-\frac{\theta_\pi}{\beta} + \frac{c_w\Qstar}{\beta} +h_\pi. \label{eq:g_qstar_nius_1}
\end{align}
The last equality follows from~\eqref{eq:g_q_beyond_Qstar}. 
Following a similar approach as~\eqref{eq:g_qstar_nius_1}, we establish a recursive relationship between $g(Q)$ and $h$ for $Q< Q^{\ast}$.  
          \begin{align}
          g(Q)&=\frac{c_w}{\beta}\sum_{l=Q+1}^{Q^{\ast}} l-\left(Q^{\ast}-Q\right)\frac{\theta_\pi}{\beta}+h_{\pi} \nonumber\\
          &=\frac{c_w(Q^{\ast}-Q)(Q^{\ast}+Q+1)}{2\beta}-\left(Q^{\ast}-Q\right)\frac{\theta_\pi}{\beta}+h_\pi. \label{eq:value_q__tau}
      \end{align}
      Hence from~\eqref{eq:value_q__tau}, \begin{align}
          g(0)&=\frac{c_w\Qstar(\Qstar+1)}{2\beta}-\frac{\Qstar\theta_\pi}{\beta}+h_\pi\nonumber\\
          g(0)&\stackrel{(a)}{=}\frac{c_w\Qstar(\Qstar+1)}{2\beta}-\frac{\Qstar\theta_\pi}{\beta}-\frac{\theta_\pi}{\beta}+c_f+g(0)\nonumber\\
          \Rightarrow& \frac{c_w\Qstar(\Qstar+1)}{2\beta}-\frac{(\Qstar+1)\theta_\pi}{\beta}+c_f=0 \label{eq:first_q}
      \end{align}
      The equality~$(a)$ is obtained by replacing the value of $h_\pi$ from~\eqref{eq:g_q_beyond_Qstar}.
According to the policy $\pi$, $\Qstar$ will be 
\begin{align}
          &\min\{Q:\frac{\theta_\pi}{\beta}+\frac{c_w(Q+1)}{\beta}+g(Q+1)\geq h_{\pi}\} \nonumber\\
           &\Rightarrow \min\{Q:\frac{\theta_\pi}{\beta}+\frac{c_w(Q+1)}{\beta}+h_{\pi}\stackrel{(b)}\geq h_{\pi}\} \nonumber\\
          &\Rightarrow \min\{Q:Q\geq \frac{{\theta_\pi}}{c_w}-1\}\nonumber\\
          &\Rightarrow \Qstar=\left\lfloor\frac{\theta_\pi}{c_w}\right\rfloor.\label{eq:theta_and_Q}
      \end{align}
The equality~$(b)$ is obtained from the fact that for $Q\geq\Qstar$, $g(Q)=h_{\pi}$.
We can obtain the value of $\Qstar$ by solving~\eqref{eq:first_q} and~\eqref{eq:theta_and_Q}.

In the following, we show that $h_\pi(Q,\tau)=g(Q)$, $\theta_\pi$ and $\Qstar$ satisfy the Bellman's equation~\eqref{bellman_eqn_with_queue_infinite_capacity} for $\tau>\taustar$, in particular $h(Q,\tau)=g(Q)$, $\theta=\theta_\pi$ and \begin{equation}Q^\ast=\left\lfloor\frac{\theta}{c_w}\right\rfloor\text{ (from~\eqref{eq:theta_and_Q})}\label{eq:opt_theta_vs_Q}\end{equation}satisfy~\eqref{bellman_eqn_with_queue_infinite_capacity}. This further implies we have to show for $Q<\Qstar$
\begin{equation}-\frac{\theta}{\beta}+\frac{c_w(Q+1)}{\beta}+g(Q+1)\leq-\frac{\theta}{\beta}+c_f+g(0)=h_\pi\label{eq:first_cond}\end{equation}
 and for $Q\geq\Qstar$,
\begin{equation}-\frac{\theta}{\beta}+\frac{c_w(Q+1)}{\beta}+g(Q+1)\geq-\frac{\theta}{\beta}+c_f+g(0)=h_\pi\label{eq:2nd_cond}\end{equation}
For $Q=\Qstar-1$ the left hand side of~\eqref{eq:first_cond} becomes \begin{align*}
     -\frac{\theta}{\beta}+\frac{c_w\Qstar}{\beta}+g(\Qstar)\stackrel{(a)}{=} -\frac{\theta}{\beta}+\frac{c_w\Qstar}{\beta}+h_\pi\leq h_{\pi}
 \end{align*}
 where the equality~$(a)$ follows from~\eqref{eq:g_q_beyond_Qstar} and the last inequality follows from the fact that $\Qstar=\left\lfloor\frac{\theta}{c_w}\right\rfloor$ from~\eqref{eq:opt_theta_vs_Q}. 
 
Let $Q<\Qstar-1$ then the left hand side of~\eqref{eq:first_cond} becomes\begin{align*}
    &-\frac{\theta}{\beta}+\frac{c_w(Q+1)}{\beta}+g(Q+1)\nonumber\\
    &=-\frac{\theta}{\beta}+\frac{c_w(Q+1)}{\beta}+\frac{c_w(Q^{\ast}-Q-1)(Q^{\ast}+Q+2)}{2\beta}\nonumber\\
    &\quad-\left(Q^{\ast}-Q-1\right)\frac{\theta_\pi}{\beta}+h_\pi\nonumber\\
    &\stackrel{(a)}{\leq}c_w(Q^{\ast}-Q-1)\frac{(\Qstar+Q+2-2\Qstar-2)}{2\beta}+h_\pi\nonumber\\
    &=c_w(Q^{\ast}-Q-1)\frac{(Q-\Qstar)}{2\beta}+h_\pi
    <h_\pi.
\end{align*}
We obtain the first equality by replacing the value of $g(Q)$ from~\eqref{eq:value_q__tau}. We obtain the inequality~$(a)$ by replacing $\theta\geq c_w\Qstar$ since $\Qstar=\left\lfloor\frac{\theta}{c_w}\right\rfloor$ from~\eqref{eq:opt_theta_vs_Q}. The last equality follows from the fact that $Q<\Qstar-1$.

In the following we will show for $Q\geq\Qstar$ ~\eqref{eq:2nd_cond} is satisfied. The left hand side of~\eqref{eq:2nd_cond} becomes
\begin{align*}
    -\frac{\theta}{\beta}+\frac{c_w(Q+1)}{\beta}+g(Q+1)&\stackrel{(a)}{=}-\frac{\theta}{\beta}+\frac{c_w(Q+1)}{\beta}+h_{\pi}\stackrel{(b)}\nonumber\\
    &\geq h_\pi.
\end{align*}
The equality~$(a)$ follows from~\eqref{eq:g_q_beyond_Qstar}. The equality~$(b)$ follows from the fact that $\Qstar-1\leq\frac{\theta}{c_w}\leq\Qstar+1$ since $\Qstar=\left\lfloor\frac{\theta}{c_w}\right\rfloor$ from~\eqref{eq:opt_theta_vs_Q}. 
Hence, we showed $\pi(Q,\tau)=\pi^{\ast}(Q, \tau)$ for $\tau>\taustar$ and we have designed $\pi$ in a way such that $\pi(q,\tau)=\pi^{\ast}(q,\tau)$ for $\tau\leq\taustar$. 
\end{IEEEproof}
\begin{lemma}
   There exist unique solutions to Equations~\eqref{eq:first_eq} and~\eqref{eq:2nd_eq}. 
\end{lemma}
\begin{IEEEproof}
After replacing~\eqref{eq:2nd_eq} in~\eqref{eq:first_eq}
    \begin{align}
    \frac{c_w\left\lfloor\frac{\theta}{c_w}\right\rfloor\left(\left\lfloor\frac{\theta}{c_w}\right\rfloor+1\right)}{2\beta}-\frac{\left(\left\lfloor\frac{\theta}{c_w}\right\rfloor+1\right)\theta}{\beta}+c_f=0.
    \end{align}
    We will show that the above equation has a unique solution.
    Let us define \begin{equation}f(\theta)\coloneqq \frac{c_w\left\lfloor\frac{\theta}{c_w}\right\rfloor\left(\left\lfloor\frac{\theta}{c_w}\right\rfloor+1\right)}{2\beta}-\frac{\left(\left\lfloor\frac{\theta}{c_w}\right\rfloor+1\right)\theta}{\beta}+c_f.\label{def:g_theta}\end{equation}
    Note that, $g(0)=c_f>0$. Now will show $f(\theta)$ is piece wise linear and decreasing in $\theta$. 
    Let $n\geq 0$, the slope of $f(\theta)$ for $\theta$ in $[nc_w \,\,(n+1)c_w)$ is $-\frac{n+1}{\beta}$ as the value of $f(\theta)$ is the following. 
    \[
       f(\theta)=\frac{c_wn(n+1)}{2\beta}-\frac{(n+1)\theta}{\beta}+c.  \]
       Hence $f(\theta)$ is linearly decreasing between $nc_w$ and $(n+1)c_w$.
       Now we will show that \[\lim_{\epsilon\to 0}\vert f((n+\epsilon)c_w)-f((n-\epsilon)c_w)\vert=0\]
       From~\eqref{def:g_theta},\[f((n+\epsilon)c_w)=\frac{c_w(n+1)(-n-2\epsilon)}{2\beta}+c.\]
       and \[f((n-\epsilon)c_w)=\frac{-c_w(n+1+2\epsilon)n}{2\beta}+c.\]
       Hence \[\lim_{\epsilon\to 0}\vert f((n+\epsilon)c_w)-f((n-\epsilon)c_w)\vert=0\]
      Hence,  $f(\theta)$ is continuous in $\theta$. 
      Since $f(0)>0$ and $f(\theta)$ is continuous,  piecewise linear, and decreasing in $\theta$, there exist  unique $\theta$ and $\Qstar=\left\lfloor\frac{ \theta}{c_w} \right\rfloor$ as solutions to~\eqref{eq:first_eq} and~\eqref{eq:2nd_eq}. 
\end{IEEEproof}
  In the next lemma we obtain the values of $Q^{\ast}$,$\taustar$ and $\theta$.
  \begin{lemma}
     \begin{enumerate}
         \item $h(q,\tau)=$ \[\begin{cases}
              c_a\lambda\tau-\frac{\beta c_a\lambda\tau^2}{2}+\theta \tau+h-c_f, \text{ for }\tau\leq \taustar, \nonumber\\
              \frac{(Q^{\ast}-q)(Q^{\ast}+q+1)c_w}{2\beta}-\left(Q^{\ast}-q\right)\frac{\theta}{\beta}+h, \nonumber\\\quad \,\text{ for }q< Q^{\ast} \text{ and }\tau>\taustar, \nonumber\\
              h, \text{ for }q\geq Q^{\ast}. \nonumber
          \end{cases}\]
          \item $\theta=\beta c_a\lambda\taustar$.
          \item $Q^{\ast}$ and $\taustar$ are solutions to the following equations:\begin{align*}
          &\tau=-\frac{(Q+1)}{\beta}+\frac{1}{\beta}\sqrt{(Q+1)^2+\frac{2\beta c_f}{c_a\lambda}+\frac{Q(Q+1)}{c_a\lambda}} \\
           &Q=\left\lfloor \frac{\beta c_a\lambda\tau}{c_w}\right\rfloor.    \end{align*}
     \end{enumerate}
  \end{lemma}
  \begin{IEEEproof}
    From the definition of $h$ and  Lemma~\ref{lemma:threshold_policy} for $\tau\geq\taustar$, we have \begin{equation}
        h= \left(Q^{\ast},\tau\right)\label{eq:value_at_qstar_is_h}
    \end{equation}
      \begin{align}
          h\left(Q^{\ast}-1,\tau\right)&=-\frac{\theta}{\beta}+\frac{c_w}{\beta}Q^{\ast}+\int_{0}^{\infty}\beta e^{-\beta t}h(Q^{\ast},t+\tau)dt\nonumber\\
          &=-\frac{\theta}{\beta}+\frac{c_w}{\beta}Q^{\ast}+h.
          \end{align}
          The last equality is obtained by replacing  $h\left(Q^{\ast},\tau\right)=h$ from~\eqref{eq:value_at_qstar_is_h}. Following a similar approach as before, we establish a recursive relationship between $h(q,\tau)$ and $h$ for $q\leq Q^{\ast}$ and $\tau\geq\taustar$.  
          \begin{align}
          &h(q,\tau)=\frac{c_w}{\beta}\sum_{l=q+1}^{Q^{\ast}} l-\left(Q^{\ast}-q\right)\frac{\theta}{\beta}+h \nonumber\\
          &=\frac{(Q^{\ast}-q)(Q^{\ast}+q+1)c_w}{2\beta}-\left(Q^{\ast}-q\right)\frac{\theta}{\beta}+h \label{eq:value_q__tau}
      \end{align}
In the following, we obtain the value of $\theta$. We define $\bar{h}:=h(0,\taustar)$ and obtain the value of $\bar{h}$ from~\eqref{eq:value_q__tau} as follows:
\begin{equation}
   \bar{h}=\frac{Q^{\ast}(Q^{\ast}+1)c_w}{2\beta}-\frac{Q^{\ast}\theta}{\beta}+h. \label{eq:value_of_hbar}
\end{equation} For $\tau\geq\taustar$, $h(q,\tau)$ only depends upon $q$ and hence, $h(0,\tau)=\bar{h}$ for $\tau\geq\taustar$. Hence from lemma~\ref{lemma:threshold_policy}, \begin{align}
  \bar{h}=h(0,\taustar)&=c_a\lambda\taustar-\frac{\theta}{\beta}+\int_{0}^{\infty}\beta e^{-\beta t}h(0,t+\taustar)dt \nonumber\\
  \implies \bar{h} &\stackrel{(a)}=c_a\lambda\taustar-\frac{\theta}{\beta}+\bar{h} \nonumber\\
  \implies \theta=\beta c_a\lambda\taustar.\label{eq:value_of_theta_with_waiting_cost}
\end{align}

Moreover, for $\tau\leq\taustar$,
      \begin{align}
          h(0,\tau)&=-\frac{\theta}{\beta}+c_a\lambda\tau+\int_{0}^{\infty}\beta e^{-\beta t}h(0,t+\tau)dt\nonumber\\
          &=-\frac{\theta}{\beta}+c_a\lambda\tau+e^{\beta \tau}\int_{\tau}^{\infty}\beta e^{-\beta t}h(0,t)dt \label{eq:for_replacement}.
      \end{align}
      The last equality is obtained by change in variables.After taking the derivative w.r.t. $\tau$, we obtain
      \begin{align}
          \dot{h}(0,\tau)&=c_a\lambda+\beta e^{\beta \tau}\int_{\tau}^{\infty}\beta e^{-\beta t}h(0,t)dt-\beta h(0,\tau)\nonumber\\
          &\stackrel{(a)}=c_a\lambda+\beta(h(0,\tau)+\frac{\theta}{\beta}-c_a\lambda\tau)-\beta h(0,\tau)\nonumber\\
          &=c_a\lambda-\beta c_a\lambda\tau+\theta \label{eq:derivative_queue_of_h_0_tau}
      \end{align}
      The equation~$(a)$ is obtained after replacing $e^{\beta \tau}\int_{\tau}^{\infty}\beta e^{-\beta t}h(0,t)dt$  from~\eqref{eq:for_replacement}. 
      After solving the differential equation~\eqref{eq:derivative_queue_of_h_0_tau}, we obtain
      \begin{equation}
          h(0,\tau)=c_a\lambda\tau-\frac{\beta c_a\lambda\tau^2}{2}+\theta \tau+C_1\label{eq:quadratic_eq_with_c_1}
      \end{equation}
      where $C_1$ is an integrating constant. In the following, we obtain the value of $C_1$. At $\tau=0$, $h(0,0,0)=h-c_f$, recall that \[h=c_f-\frac{\theta}{\beta}+\int_{0}^{\infty}\beta e^{-\beta t}h(0,t)dt \text{ from~\eqref{bellman_eqn_with_queue_infinite_capacity}}.\] Moreover, $C_1=h(0,0)=h-c_f$. We replace the value of $C_1$ in~\eqref{eq:quadratic_eq_with_c_1} and obtain for $\tau\leq\taustar$,\begin{equation}
          h(0,\tau)=c_a\lambda\tau-\frac{\beta c_a\lambda\tau^2}{2}+\theta \tau+h-c_f. \label{eq:quadratic_of_h_0_tau}
      \end{equation}
      As we defined earlier,
          \[ \bar{h}=h(0,\taustar)=c_a\lambda\taustar-\frac{\beta c_a\lambda{\taustar}^2}{2}+\theta \taustar+h-c_f. \] 
      From~\eqref{eq:value_of_hbar}, \begin{align}
          \frac{Q^{\ast}(Q^{\ast}+1)c_w}{2\beta}-\frac{Q^{\ast}\theta}{\beta}+h&=c_a\lambda\taustar-\frac{\beta c_a\lambda{\taustar}^2}{2}+\theta \taustar\nonumber\\
          &\quad+h-c_f \nonumber\\
          \Rightarrow c_a\lambda\taustar-\frac{\beta c_a\lambda{\taustar}^2}{2}+\theta \taustar+&\frac{Q^{\ast}\theta}{\beta}-\frac{Q^{\ast}(Q^{\ast}+1)c_w}{2\beta}=c_f\nonumber\\
            \Rightarrow (Q^{\ast}+1)c_a\lambda\taustar+\frac{\beta c_a\lambda{\taustar}^2}{2}&-\frac{Q^{\ast}(Q^{\ast}+1)c_w}{2\beta}-c_f\stackrel{(a)}=0. \label{eq:between_taub_and_Qstar}\end{align}
      The equality~$(a)$ is obtained by replacing the value of $\theta= \beta c_a\lambda\taustar$  from~\eqref{eq:value_of_theta_with_waiting_cost}. After solving the quadratic equation~\eqref{eq:between_taub_and_Qstar}, we obtain
      \begin{align}
          \taustar=-\frac{(Q^{\ast}+1)}{\beta}+\frac{1}{\beta}\sqrt{(Q^{\ast}+1)^2+\frac{2\beta c_f}{c_a\lambda}+\frac{Q^{\ast}(Q^{\ast}+1)c_w}{c_a\lambda}} \label{eq1:between_taub_and_Qstar}
      \end{align} This equation provides a relation between $ Q^{\ast}$ and $\taustar$. To obtain the values of $ Q^{\ast}$ and $\taustar$ we require another equation and we will obtain it in the following.
      
       Recall from Lemma~\ref{lemma:threshold_policy} $Q^{\ast}=\lfloor \frac{\theta}{c_w}\rfloor$. After  replacing the value of $\theta= \beta c_a\lambda\taustar$  from~\eqref{eq:value_of_theta_with_waiting_cost} we obtain, 
      \begin{equation}
          Q^{\ast}=\left\lfloor \frac{\beta c_a\lambda\taustar}{c_w}\right\rfloor \label{eq2:between_taub_and_Qstar}
      \end{equation}
\remove{
      Since $h(q,\tau)=h$ for $\tau\geq\taustar$ and $q\geq Q^{\ast}$ from~\eqref{eq:value_at_qstar_is_h} and for $\tau\geq\taustar$,
      $h(Q^{\ast},\tau)=\min\{-\frac{\theta}{\beta}+\frac{c_w}{\beta}(Q^{\ast}+1)+\int_{0}^{\infty}\beta e^{-\beta t}h(Q^{\ast}+1,t+\tau)dt,h\}$, we have for $q\geq Q^{\ast}$ and $\tau\geq\taustar$\begin{align}
          &\frac{\theta}{\beta}+\frac{c_w}{\beta}(q+1)+\int_{0}^{\infty}\beta e^{-\beta t}h(q+1,t+\tau)dt\geq h \nonumber\\
           \implies& \frac{\theta}{\beta}+\frac{c_w}{\beta}(q+1)+h\stackrel{(a)}\geq h \nonumber\\
          \implies& q\geq \frac{\theta}{c_w}-1
      \end{align}
      The equality~$(a)$ is obtained from lemma~\ref{lemma:threshold_policy}. Hence, $Q^{\ast}=\min_{\{q\in\mathbb{N}\}}\{q\geq \frac{\theta}{c_w}-1\}$. }
      \end{IEEEproof}
      \begin{lemma}\label{lemma:uniqueness_of_qstar_and_taustar}
          There exists a unique solution to the following equations:
         \begin{align}
          &\tau=-\frac{(Q+1)}{\beta}+\frac{1}{\beta}\sqrt{(Q+1)^2+\frac{2\beta c_f}{c_a\lambda}+\frac{Q(Q+1)}{c_a\lambda}} \label{eq:1_for_tau}\\
           &Q=\left\lfloor \frac{\beta c_a\lambda\tau}{c_w}\right\rfloor \label{eq:2_for_tau} 
      \end{align}
      \end{lemma}
     \begin{IEEEproof}
         We will first show that there exists a solution to these equations. We can rewrite~\eqref{eq:2_for_tau} as $\tau=\frac{c_wQ}{\beta c_a\lambda}$, where $Q\in \mathbb{N}$. We compute the ratio of $\tau$ from~\eqref{eq:1_for_tau} and $\tau=\frac{c_wQ}{\beta c_a\lambda}$ and let us define the ration as $r(Q)$. Hence, $r(Q)=$\[\frac{\lambda c_a(Q+1)}{Q c_w}+\sqrt{\frac{(Q+1)^2\lambda^2c_a^2}{Q^2c_w^2}+\frac{2\beta c_fc_a\lambda}{Q^2c_w^2}+\frac{(Q+1)c_a\lambda}{Qc_w}}.\]
         We take limit from the both sides and obtain
         \begin{align}
             \lim_{Q\to\infty}r(Q)&=-\frac{\lambda c_a}{c_w}+\sqrt{(\frac{\lambda c_a}{c_w})^2+\frac{\lambda c_a}{c_w}}\nonumber\\
             &=\frac{\lambda c_a}{c_w}(\sqrt{1+\frac{c_w}{\lambda c_a}}-1)\nonumber
         \end{align}
         Hence, $\lim_{Q\to\infty}r(Q)<1$ since $\sqrt{1+\frac{c_w}{\lambda c_a}}<1+\frac{c_w}{\lambda c_a}.$ 
         Hence for large enough $Q$, the line $\tau=\frac{c_wQ}{\beta c_a\lambda}$ lies above the curve in~\eqref{eq:1_for_tau} and $Q=0$, the line $\tau=\frac{c_wQ}{\beta c_a\lambda}$ lies below the curve in~\eqref{eq:1_for_tau}. Hence there exists a solution to~\eqref{eq:1_for_tau} and~\eqref{eq:2_for_tau}. 

         The uniqueness follows from the fact that the optimal cost, $\theta=\beta c_a\lambda\taustar$. 
     \end{IEEEproof}
  \subsection{Proof of Theorem~\ref{thm:fin_cap_wait}}
  \label{apndx:prf_main_thm}
We state the monotonicity properties of the relative cost functions in the following lemma. These properties are used in derivation of the optimal policy. 
\begin{lemma}
  \begin{enumerate}[(a)] \item $h(0,\tau,1,1)$ and $h(0,\tau,1,0)$ are non-decreasing in $\tau$.
  $h(Q,0,1)$ and $h(Q,0,0)$ are non-decreasing in $Q$. 
  \item For a given $r\geq 0$, $L_r(\tau)$ is non-decreasing in $\tau$.
  \end{enumerate}
  \end{lemma}
  \begin{IEEEproof}
   Part~$(a)$ has a standard proof using relative value iteration~\cite{bertsekas2011dynamic}. Since the integrand in the definition of $L_r(\tau)$ is  non-decreasing in $\tau$, so is   $L_r(\tau)$. This establishes Part~$(b)$. 
  \end{IEEEproof}
 Using~\eqref{eqn:bellman_hq00}, we can rewrite the the Bellman's equations as follows.
\begingroup
\allowdisplaybreaks
\begin{align}
 h(0,\tau,1,1) = &\min \left\{
     c_a\lambda \tau +\frac{C_h-\theta}{\beta}+\Lotau,\right.\nonumber\\
     &\left. c_f+\frac{C_h-\theta}{\beta} +\Lot,h(1,0,0),\right.\nonumber \\
      &c_f+h(0,0,0),c_a\lambda\tau+h(0,0,0) \bigg\} \label{eqn:bellman_h0tau11}\\
      h(0,\tau,1,0) =&\min \left\{
     \frac{C_h-\theta}{\beta}+\Lotau, h(0,0,0)\right\}\label{eqn:bellman_h0tau10}\\
 h(Q,0,1) =&\min \left\{
c_f+\frac{C_h-\theta}{\beta}+\Lot,\right.\nonumber\\
&h(Q+1,0,0),c_f+h(0,0,0)\bigg\}. \label{eqn:bellman_hq01}
  \end{align}
\endgroup 
Note that, from~\eqref{eqn:bellman_hq00} we can get the following relations between the cost function $h(Q,0,0)$ and $h(Q,0,1)$. 
\begin{equation}
    h(Q,0,0)=\frac{Qc_w-\theta}{p\beta}+h(Q,0,1). \label{eqn:rel_hq00_and_hq01}
\end{equation}
Based on the Bellman's equation~\eqref{eqn:bellman_h0tau10}
we can divide the optimal actions in the state in $(0,\tau,1,0)$ in two cases as follows. \begin{enumerate}
    \item \label{case_1_queue}$\chthetabybeta+\Lot\geq \hzero$. 
   This implies the optimal action is to evict for all $\tau\geq0$ in states $(0,\tau,1,0)$ and $(0,\tau,1,1)$ from~\eqref{eqn:bellman_h0tau11} and~\eqref{eqn:bellman_h0tau10}. So, given any initial state, in steady state the underlying Markov chain will have only the states in $\{(Q,0,0),(Q,0,1)\}$. Hence, the states $\{(0,\tau,1,1),(0,\tau,1,0), \tau\geq 0  \}$ are transient. In this case, from~\eqref{eqn:bellman_hq01},
    \begin{equation}
        \hqzeroone=\min\{c_f+\hzero,h(Q+1,0,0)\}. 
    \end{equation}
Furthermore, from~\eqref{eqn:rel_hq00_and_hq01}, \[h(Q+1,0,0)=\frac{(Q+1)c_w-\theta}{p\beta}+h(Q+1,0,1).\] Since $c_f+\hzero$ is a constant and $h(Q,0,0)$ is increasing $Q$ from~\eqref{eqn:rel_hq00_and_hq01}, there exists a $\Qhat\geq 0$ such that $h(Q,0,0)\geq c_f+\hzero$ for all $Q\geq \Qhat$. Hence, \begin{align}
    \hqzeroone=\begin{cases}
        \frac{(Q+1)c_w-\theta}{p\beta}+h(Q+1,0,1) \text{ for }Q<\Qhat,\\
        c_f+\hzero \text{ for }Q\geq\Qhat.\label{hQ01_under_redc_st}
    \end{cases}
\end{align} 
Hence, $h(\Qhat,0,1)=c_f+\hzero$ and
\[h(\Qhat-1,0,1)= \frac{\Qhat c_w-\theta}{p\beta}+h(\Qhat,0,1) \]
Repeating this for $Q\leq\Qhat$, we obtain
\begin{align}
    \hqzeroone=\frac{c_w}{p\beta}\sum_{l=Q+1}^{\Qhat}l-\frac{(\Qhat-Q)\theta}{p\beta}+c_f+\hzero.\label{eq:h_q01_via_h000}
\end{align}
Furthermore, after setting $Q=0$, we obtain
\begin{equation}
    \hzeroone=\frac{c_w \Qhat (\Qhat+1)}{2p\beta}-\frac{\Qhat\theta}{p\beta}+c_f+\hzero. \label{eq:h_001_via_h000}
\end{equation}
From the relation between $\hzero$ and $\hzeroone$ by substituting $Q=0$ in~\eqref{eqn:rel_hq00_and_hq01}, we obtain
\begin{align}
    \hzeroone=&\frac{c_w \Qhat (\Qhat+1)}{2p\beta}-\frac{\Qhat\theta}{p\beta}+c_f\nonumber\\
    &-\frac{\theta}{p\beta}+\hzeroone \nonumber\\
    \implies \theta =&\frac{2p\beta c_f+c_w\hat{Q}(\hat{Q}+1)}{2(\hat{Q}+1)} \label{eq:theta_reduc_st}
\end{align}
By the definition of $\Qhat$,
Hence, $\Qhat\leq \frac{\theta}{c_w}$. Since
\begin{align*}
    \Qhat&=\min\left\{Q:h(Q,0,1)\geq c_f+\hzero\right\}\nonumber\\
    &=\min\bigg\{Q:\frac{(Q+1) c_w-\theta}{p\beta}+h(Q+1),0,1)\\
    &\geq c_f+\hzero\bigg\}
\end{align*}
Since $h(Q,0,1)=c_f+h(0,0,0)$ for $q\geq\Qhat$, we have \begin{equation}\Qhat=\min\{Q:Q\geq \frac{\theta}{c_w}-1\}=\left\lfloor\frac{\theta}{c_w}\right\rfloor.\label{eq:Qhat_c_w}\end{equation}
After substituting the value of $\theta$ from~\eqref{eq:theta_reduc_st}, we obtain \begin{equation}
    \hat{ Q}=\left\lfloor\frac{2p\beta c_f+ c_w\hat{Q}(\hat{Q}+1)}{2c_w(\hat{Q}+1)}\right\rfloor.\label{eq:theta_first_case}
\end{equation}
The optimal policy from~\eqref{hQ01_under_redc_st} is \[\pi^{\ast}(s)=\begin{cases}
          2 \text{ for }Q< \hat{Q}\\
          3 \text{ for }Q\geq \hat{Q}.
      \end{cases}\]
We will derive the values of $C_h$ for which this case (Case~\ref{case_1_queue}) is satisfied after discussing the next case.  
\item \label{case_2_queue}$\chthetabybeta+\Lot< \hzero$. Define $\taub$ such that \[\taub \coloneqq \min \left\{\tau>0:\frac{C_h-\theta}{\beta}+\Lotau
    \geq h(0,0,0)\right \}.\]
    Then from~\eqref{eqn:bellman_h0tau10},
     \begin{align}
        h(0,\tau,1,0)&=\begin{cases}
          \frac{C_h-\theta}{\beta}+\Lotau \,\, \forall \,\tau \leq \taub \\
          h(0,0,0)\,\, \forall \, \tau > \taub. 
          \label{eq:h_0tau_1_0_full_st}
        \end{cases} 
        \end{align}
        Let us define: $\hat{h}\coloneqq \chthetabybeta+c_f+\Lot$. From the condition of this case (Case~\ref{case_2_queue}) $c_f+h(0,0,0)>\hat{h}$ and therefore from~\eqref{eqn:bellman_hq01},
        \begin{equation}
            \hqzeroone=\min\{h(Q+1,0,0),\hat{h}\} \label{eq:h_0_0_1_full_st}
        \end{equation}
        Since $\chthetabybeta+\Lot< \hzero$ in this case, we can further show that 
        \begin{align}
 h(0,\tau,1,1) = &\min \left\{
     c_a\lambda \tau +\frac{C_h-\theta}{\beta}+\Lotau,\hat{h},\right.\nonumber\\
     & \quad \quad h(1,0,0), c_a\lambda\tau+h(0,0,0)\bigg\} \nonumber\\
     =&\min \left\{
     c_a\lambda \tau +\frac{C_h-\theta}{\beta}+\Lotau,h(0,0,1),\right.\nonumber\\
     &\quad \quad c_a\lambda\tau+h(0,0,0)\bigg\} \label{eqn:bellman_h0tau11_full_state}
      \end{align}
      The last equality is achieved by substituting \[\hzeroone=\min\{h(1,0,0),\hat{h}\}\] from~\eqref{eq:h_0_0_1_full_st}. 

      Since $h(0,0,1)$ is a constant, there exists a $\taustar>0$ such that \[c_a\lambda\taustar+\chthetabybeta+L_0(\taustar)\geq h(0,0,1).\] Using this fact, we can further divide case~\ref{case_2_queue} into two further subcases as follows.
      \paragraph{} $\taustar\leq\taub.$ By the definition of $\taub$  \[\frac{C_h-\theta}{\beta}+\Lotau <h(0,0,0)\text{ for }\tau\leq\taub.\] By the definition of $\taustar$, \[h(0,0,1)\leq c_a\lambda\tau+\frac{C_h-\theta}{\beta}+\Lotau \text{ for }\tau\geq\taustar.\]  Since $\taustar\leq\taub$, we have
          \begin{align}
 h(0,\tau,1,1) =
     \begin{cases}
          c_a\lambda \tau +\frac{C_h-\theta}{\beta}+\Lotau \text{ for }\tau\leq\taustar \\
          h(0,0,1) 
     \end{cases} \label{eqn:bellman_h0tau11_subcase_1}
 \end{align} 
 In the following we obtain the value of $C_h$ for which the the following condition is satisfied.
 \begin{equation}
     \chthetabybeta+L_0(\taub)=h(0,0,0) \label{eq:to_obtain_ch}
 \end{equation}
 Recall the definition of $L_0(\taub)$ from~\eqref{eq:lotau_queue},
 \begin{align}
     L_0(\taub)&=\int_{0}^{\infty}\beta e^{-\beta t}\Big(ph(0,t+\taub,1,1)\nonumber\\
     &+(1-p)h(0,t+\taub,1,0)\Big)dt \nonumber\\
     &=ph(0,0,1)+(1-p)h(0,0,0) \label{eq:Lotaub_subcase_1}
 \end{align}
 The last equality is obtained by substituting $h(0,\tau,1,1)=\hzeroone$ and $h(0,\tau,1,0)=\hzero$ for $\tau\geq\taub$ from~\eqref{eqn:bellman_h0tau11_subcase_1} and~\eqref{eq:h_0tau_1_0_full_st}, respectively. We further substitute the value of $\Lotau$ from~\eqref{eq:Lotaub_subcase_1} in~\eqref{eq:to_obtain_ch} and obtain, 
 \begin{align}
 &\chthetabybeta+ph(0,0,1)+(1-p)h(0,0,0)=h(0,0,0)\nonumber\\
     &\implies C_h=0 \label{eq:C_h_0}
 \end{align}
 The last equality is obtained by substituting the value of $h(0,0,0)=-\frac{\theta}{p\beta}+h(0,0,1)$ from~\eqref{eqn:rel_hq00_and_hq01}.

 Furthermore, by the definition of $\taustar$, from~\eqref{eqn:bellman_h0tau11_subcase_1}, we obtain
\begin{align}
    &c_a\lambda\taustar+\chthetabybeta+L_0(0,\taustar)=h(0,0,1)\nonumber\\
    &c_a\lambda\taustar-\frac{\theta}{p\beta}+h(0,0,0)\stackrel{(a)}=h(0,0,1)\nonumber\\
    &\theta=p\beta c_a\lambda\taustar \label{eq:value_of_theta_subcase_1}
\end{align}
where the equality~$(a)$ follows by replacing $C_h=0$ from~\eqref{eq:C_h_0} and replacing $L_0(0,\taustar)=h(0,0,0)$ from~\eqref{eq:h_0tau_1_0_full_st} as $\taub=\taustar$. The equality~$(b)$ follows from~\eqref{eqn:rel_hq00_and_hq01}.

  Since $h(Q,0,1)$ is increasing in $Q$, there exists a $\Qstar>0$ such that 
\[\Qstar=\min\{Q:\frac{(Q+1)c_w-\theta}{\beta}+h(Q+1,0,1)\geq h\}.\] Hence, from~\eqref{eq:h_0_0_1_full_st}  
\begin{align}
    \hqzeroone=\begin{cases}
        \frac{(Q+1)c_w-\theta}{p\beta}+h(Q+1,0,1) \text{ for }Q<\Qstar,\\
        \hat{h} \text{ for }Q\geq\Qstar.\label{hQ01_under_full_st}
    \end{cases}
\end{align}
After following similar steps in obtaining~\eqref{eq:h_q01_via_h000},~\eqref{eq:h_001_via_h000} and~\eqref{eq:Qhat_c_w}, respectively, we obtain

\begin{align}
    \hqzeroone&=\frac{c_w}{p\beta}\sum_{l=Q+1}^{\Qstar}l-\frac{(\Qstar-Q)\theta}{p\beta}+\hat{h},\nonumber\\
    \hzeroone&=\frac{c_w \Qstar (\Qstar+1)}{2p\beta}-\frac{\Qstar\theta}{p\beta}+\hat{h} \label{eq:h_001_via_h000_subcase_1}, \text{ and},\\ \Qstar&=\left\lfloor\frac{\theta}{c_w}\right\rfloor.\label{eq:Qstar_theta}
\end{align}
We note that if $Q^\ast=0$, then $h(0,0,1)=\hat{h}$, and the state with $(Q,0,1)$ 
for $Q>0$ is never visited under optimal actions. 
After substituting the value of $\theta$ in~\eqref{eq:Qstar_theta} from~\eqref{eq:value_of_theta_subcase_1} we obtain 
\begin{equation}
    \Qstar=\left\lfloor\frac{p\beta c_a\lambda \taustar}{c_w}\right\rfloor. \label{eq:floor_taustar_qstar}
\end{equation}
This provides a relation between $\Qstar$ and $\taustar$. In the following, we derive another relation between $\Qstar$ and $\taustar$. For this we need following Lemma. 

    \begin{table*}[h]
    \centering
\caption{States and Optimal actions for case $2a)$}
\label{table:case2optimalaction}
\setlength\tabcolsep{2pt}
\begin{tabular}{|c|c|c|}
\hline
    \multicolumn{3}{|c|}{ Optimal actions} \\
  \hline 
    $s=(0,\tau,1,1)$ &  $s=(0,\tau,1,0)$ & $s=(Q,0,1)$\\ 
     \hline
      $\pi^{\ast}(s)=\begin{cases}
          0\text{ for } \tau \leq \taustar \\
          2\text{ for } \tau > \taustar,\, Q^{\ast}>0\\
          1\text{ for }\tau>\taustar,\,Q^{\ast}=0
      \end{cases}$ & $\pi^{\ast}(s)=0$ & $\pi^{\ast}(s)=\begin{cases}
          2 \text{ for }Q< Q^{\ast}\\
          1 \text{ for }Q\geq Q^{\ast}
      \end{cases}$\\
     \hline
\end{tabular}
\end{table*}

In the  following Lemma we provide  the expressions for $\htauone$ and $\htauzero.$
\begin{lemma}
\label{lemma:subcase_2_values}
    $(a)$   \begin{align}&\htauone=\nonumber\\
        &\begin{cases}
            c_a\lambda \tau-p \beta c_a\lambda\frac{\tau^2}{2} +\theta \tau+\hat{h}-c_f \text{ for $\tau \leq \taustar$}\\
            h(0,0,1) \text{ for $\tau > \taustar$}
        \end{cases}\label{eq:final_h(tau,1,1)} \end{align} 
          \begin{align}&\text{and }\htauzero=\nonumber\\
        &\begin{cases}-p \beta c_a\lambda\frac{\tau^2}{2} +\theta \tau+h(0,0,1)-c_f \text{ for $\tau \leq \taub$}\\\
        h(0,0,0) \text{ for } \tau>\taub \end{cases}\label{eq:final_h(tau,1,0)}\end{align}
        $(b)$ $\taustar=\taub.$
\end{lemma}
\begin{IEEEproof}
    See Appendix~\ref{appendix:proof_of_lemma_6}. 
    \end{IEEEproof}

From ~\eqref{eq:final_h(tau,1,1)}, 
\begin{align}
    &h(0,\taustar,1,1)=h(0,0,1)\nonumber\\
    &c_a\lambda \tau-p \beta c_a\lambda\frac{{\taustar}^2}{2} +\theta \taustar+\hat{h}-c_f\stackrel{(a)}=h(0,0,1) \nonumber\\
    &c_a\lambda \taustar-p \beta c_a\lambda\frac{{\taustar}^2}{2} +\theta \taustar-c_f\stackrel{(b)}=\frac{c_w \Qstar (\Qstar+1)}{2p\beta}-\frac{\Qstar\theta}{p\beta} \nonumber\\
    & p \beta c_a\lambda\frac{{\taustar}^2}{2}+(\Qstar+1)c_a\lambda\taustar-\frac{c_w \Qstar (\Qstar+1)}{2p\beta}-c_f\stackrel{(c)}=0\label{eq:taustar_qstar}
\end{align}
where the equality~$(a)$ follows from~\eqref{eq:final_h(tau,1,1)} by substituting
\[ h(\taustar,1,1)=c_a\lambda \tau-p \beta c_a\lambda\frac{\tau^2}{2} +\theta \tau+\hat{h}-c_f.\] The equality~$(b)$ follows from~\eqref{eq:h_001_via_h000_subcase_1}. The equality~$(c)$ follows by substituting $\theta$ from~\eqref{eq:value_of_theta_subcase_1}.

The value of $\Qstar$ and $\taustar$ can be obtained by solving~\eqref{eq:floor_taustar_qstar} and~\eqref{eq:taustar_qstar}. The uniqueness follows from Lemma~\ref{lemma:uniqueness_of_qstar_and_taustar}. Finally, from~\eqref{eq:h_0tau_1_0_full_st},~\eqref{eqn:bellman_h0tau11_subcase_1}, and~\eqref{hQ01_under_full_st} we summarize the optimal actions for all states in the Table~\ref{table:case2optimalaction}.

   \paragraph{}  $\taustar>\taub$.
  Recall the definition of $\taustar$, 
  \[\taustar=\min\{\tau: c_a\lambda\tau+\chthetabybeta+L_0(\tau)\geq h(0,0,1)\}.\]
Since $\taub\leq\taustar$ the following is true for all $\tau\leq\taub$. 
\[c_a\lambda\tau+\chthetabybeta+L_0(\tau)\leq \min\{ c_a\lambda\tau+h(0,0,0),h(0,0,1)\}.\] We next compare $c_a\lambda\tau+h(0,0,0)$ and $h(0,0,1)$ for $\tau\geq\taub$.  At $\taustar$ the curve $L_0(\tau)+\frac{C_h-\theta}{\beta}+c_a\lambda\tau$ crosses $h(0,0,1)$. Since  the slope of $L_0(\tau)+\frac{C_h-\theta}{\beta}+c_a\lambda\tau$ is at least the slope of $c_a\lambda\tau+h(0,0,0)$, the latter will cross $h(0,0,1)$ at higher value than $\taustar$. Hence, there exists a $\tilde{\tau} \geq \taustar$ such that \[\tautil{=}\min\left\{\tau >0:c_a\lambda \tau {+} h(0,0,0) {=} c_f{+}\frac{C_h-\theta}{\beta}{+}\Lot\right\}.\]         
Following the above discussions, we obtain the value of $ h(0,\tau,1,1)$ as below. 
\begin{align}
    h(0,\tau,1,1)=\begin{cases}
         c_a\lambda \tau +\frac{C_h-\theta}{\beta}+\Lotau \text{ for }\tau\leq\taub \\
          c_a\lambda \tau+h(0,0,0)  \text{ for }\taub<\tau\leq \tautil\\
          h(0,0,1)  \text{ for }\tau>\tautil.
    \end{cases} \label{eqn:bellman_h0tau11_subcase_2}
\end{align}
In the following we obtain the value of $h(Q,0,1)$.
 Let $\Qbar$ be defined as follows. 
\[\Qbar\coloneqq\left\{Q:\frac{(Q+1)c_w-\theta}{\beta}+h(Q+1,0,1)\geq \hat{h}\right\}.\] Hence, from~\eqref{eq:h_0_0_1_full_st}  
\begin{align}
    \hqzeroone=\begin{cases}
        \frac{(Q+1)c_w-\theta}{p\beta}+h(Q+1,0,1) \text{ for }Q<\Qbar,\\
       \hat{h} \text{ for }Q\geq\Qbar.\label{hQ01_under_subcase_2}
    \end{cases}
\end{align} 
     We summarize the optimal actions for all states in the Table~\ref{table:case3optimalaction}. 
     \begin{table*}[h]
     \centering
\caption{States and Optimal actions for case $2b)$}
\setlength\tabcolsep{2pt}
\label{table:case3optimalaction}
\begin{tabular}{|c|c|c|c|}
\hline
    \multicolumn{3}{|c|}{Optimal actions} \\
  \hline 
   $s=(\tau,1,1)$ &  $s=(\tau,1,0)$ & $s=(0,1)$\\ 
     \hline
      $\pi^{\ast}(s)=\begin{cases}
          0\text{ for } \tau \leq \taub \\
          4\text{ for } \taub\leq\tau \leq \tautil \\
          2\text{ for } \tau>\tautil,\, \bar{Q}>0\\
          1 \text{ for } \tau>\tautil,\, \bar{Q}=0
      \end{cases}$ & $\pi^{\ast}(s)=\begin{cases}
          0  \text{ for }\tau\leq\taub \\ 
          2 \text{ for }\tau\geq \taub
      \end{cases}$ & $\pi^{\ast}(s)=\begin{cases}
          2 \text{ for }Q< \bar{Q}\\
          1 \text{ for }Q\geq \bar{Q}
      \end{cases}$\\
     \hline
\end{tabular}
\end{table*}
In the following we obtain the values of $\Qbar,\taub,\tautil$ and $\theta$.

In the following lemma we obtain the values of $\theta,h(0,\tau,1,0)$, and $h(0,0,1)$. We also the value of $C_h$ for which Case~\ref{case_2_queue}b) is satisfied. 
\begin{lemma}
\label{lemma:case_3_values_queue}
    \begin{enumerate}
    \item $\theta=c_a \tilde{\tau} \lambda p \beta$ 
        \item $h(0,\tau,1,1)=$\begin{align*}\begin{cases}
            c_a\lambda\tau{+}c_a \lambda p \beta\left(\tilde{\tau} \tau{-}\frac{\tau^2}{2}\right){-}C_h\tau{+} \hzeroone{-}c_f \text{ for } \tau {\leq} {\taub} \\
            c_a\lambda\tau+\hzero \text{ for } \taub<\tau\leq\tautil\\
            \hzeroone \,\,\text{for } \tau > \tilde{\tau} 
\end{cases}
        \end{align*}
        \item \hzeroone=\begin{align*}\hzeroone=\frac{c_w \Qbar (\Qbar+1)}{2p\beta}-\frac{\Qbar\theta}{p\beta}+\hat{h}\end{align*}
        \item $\tautil$, $\taub$ and $\Qbar$ are solutions to the following equations.  \begin{align*}
            &(\Qbar{+}1)c_a\lambda\tautil{+}p \beta c_a \lambda \left(\tautil \taub-\frac{\taub^2}{2}\right){-}\nonumber\\ &\frac{c_w \Qbar (\Qbar+1)}{2p\beta}\stackrel{(c)}{=}0\nonumber\\
            &p c_a \lambda+C_h-p c_a \lambda \left(\beta(\tautil-\taub)+e^{-\beta(\tautil-\taub)}\right)=0 \nonumber\\
            &\Qbar=\left\lfloor\frac{p\beta c_a\lambda \tautil}{c_w}\right\rfloor.
        \end{align*}
        \item $0< C_h\leq I$ and $\tautil\leq\tau^0$ where \[I=p \beta c_f-p c_a \lambda (1-e^{-\beta \tau^0})\] and \[\tau^0=\frac{2p\beta c_f+c_w\hat{Q}(\hat{Q}+1)}{2p\beta c_a\lambda(\hat{Q}+1)}.\] 
    \end{enumerate}
\end{lemma}
\begin{IEEEproof}
    Please see Appendix~\ref{appendix:proof_of_lemma_7}. 
\end{IEEEproof}
\end{enumerate}
Recall Case~\ref{case_2_queue} or \[\chthetabybeta+\Lot< h(0,0,0)\implies 0\leq C_h\leq I.\]
Since Case~\ref{case_1_queue} and Case~\ref{case_2_queue} are mutually exclusive the following is true. \[C_h>I \implies \chthetabybeta+\Lot\geq h(0,0,0) \text{ or Case~\ref{case_1_queue}}\] 
Furthermore if $\chthetabybeta+\Lot< h(0,0,0)\iff 0\leq C_h\leq I$ then $C_h>I \iff \chthetabybeta+\Lot\geq h(0,0,0)$. We already have $\chthetabybeta+\Lot< h(0,0,0)\implies 0\leq C_h\leq I$. In the following we will show that \[C_h\leq I\implies \chthetabybeta+\Lot<\hzero.\] Suppose, $C_h\leq I\implies $ Case~\ref{case_1_queue}. The optimal cost in Case~\ref{case_1_queue} from~\eqref{eq:theta_reduc_st} is \[\theta_{Case_1}=\frac{2p\beta c_f+c_w\hat{Q}(\hat{Q}+1)}{2(\hat{Q}+1)}.\] We can alternatively write $\theta=p\beta c_a\lambda \tau^0$, where \[\tau^0=\frac{2p\beta c_f+c_w\hat{Q}(\hat{Q}+1)}{2p\beta c_a\lambda(\hat{Q}+1)}.\] Note that in Case~\ref{case_2_queue} the optimal cost $\theta_{Case_2}=p\beta c_a\lambda\tautil$ from Lemma~\ref{lemma:case_3_values_queue}. Also from Lemma~\ref{lemma:case_3_values_queue}, we have $\tautil\leq\tau^0$ and hence, \[\theta_{Case_2}\leq p\beta c_a\lambda\tau^0=\theta_{Case_1}.\] Under the condition of $C_h\leq I$, the cost under Case~\ref{case_2_queue} is lesser than the cost under Case~\ref{case_1_queue}. Hence, 
\[C_h\leq I\rightarrow \chthetabybeta+\Lot<\hzero.\]
Note that, if $C_h\leq I$, then the optimal policy is as in Tables~\ref{table:case2optimalaction} and~\ref{table:case3optimalaction}. The thresholds $\Qstar$, $\Qbar$, $\taustar$,$\taub$ and $\tautil$ from Lemmas~\ref{lemma:subcase_2_values} and~\ref{lemma:case_3_values_queue}.  
 In a similar manner we can show that $C_h=0\iff$ case~\ref{case_2_queue}a) and $0<C_h\leq I \iff$ case~\ref{case_2_queue}b). 

 $C_h>I\iff \text{Case~\ref{case_1_queue}}$ the optimal cost and the threshold value of queue length $Q$ can be obtained from~\ref{eq:theta_reduc_st} and~\ref{eq:theta_first_case},respectively. 
\subsection{Proof of Lemma~\ref{lemma:subcase_2_values}}
\label{appendix:proof_of_lemma_6}
    We use change of variables to get, $\Lotau=e^{\beta \tau}L_{\tau}(0)$. The derivative of $e^{\beta \tau}L_{\tau}(0)$ w.r.t $\tau$ is as follows:
    \begin{align}
        \frac{d}{dt}\{e^{\beta \tau}L_{\tau}(0)\}{=}&\beta e^{\beta \tau}L_{\tau}(0){-}\beta\left(p\htauone{+}\right.\nonumber\\
        &\left.(1{-}p)\htauzero\right) \label{eq:derivative_queue}
    \end{align}
    From~\eqref{eqn:bellman_h0tau11_subcase_1}  we observe that for $\tau \leq \taustar$,\begin{align}
         \htauone= c_a\lambda\tau+\frac{C_h-\theta}{\beta}+\Lotau \label{eq:h(tau,1,1)_before_derivative_queue}
    \end{align}
To obtain the derivative of $\htauone$ we use~\eqref{eq:derivative_queue} and we get, 
    \begin{align}
        \dot{h}(0,\tau,1,1){=}&c_a\lambda+\beta e^{\beta \tau}L_{\tau}(0){-}\beta\left(p\htauone{+}\right.\nonumber\\
        &\left.(1{-}p)\htauzero\right)\label{eq:h(tau,1,1)_after_derivative}\end{align}
        In the following we obtain $\htauone$ for $\tau\leq \taustar$. For this we replace $e^{\beta \tau}L_{\tau}(0)$ from~\eqref{eq:h(tau,1,1)_before_derivative_queue} and note that since $\taustar\leq\taub$ $\htauzero=\htauone-c_a\lambda \tau$ from~\eqref{eqn:bellman_h0tau11_subcase_1} and~\eqref{eq:h_0tau_1_0_full_st}  and hence we obtain, $\dot{h}(\tau,1,1)$
        \begin{align}
        =&c_a\lambda-\beta \Big(\htauone-(1-p)c_a\lambda\tau-\htauone{+}c_a\lambda\tau \nonumber\\
        &\left.+\frac{C_h-\theta}{\beta}\right) \nonumber\\
        =&c_a\lambda-p \beta c_a\lambda\tau +\theta-C_h \nonumber
    \end{align}
    We put $C_h=0$ from~\eqref{eq:C_h_0} and the solution of the above differential equation is,  \begin{align}
        \htauone=&c_a\lambda \tau-p \beta c_a\lambda\frac{\tau^2}{2} +\theta \tau +C_1 \text{ for } \tau \leq \taustar \label{eq:for_replacing_C_1}
        \end{align}
        Here, $C_1$ is an integration constant. To find the value of $C_1$, we use the value of $\htauone$ at $\tau=0$.
        \begin{align*}
       C_1= h(0,0,1,1)\stackrel{(a)}=\frac{C_h-\theta}{\beta}+\Lot\stackrel{(b)}=\hat{h}-c_f
        \end{align*}
where~$(a)$ follows from~\eqref{eqn:bellman_h0tau11_subcase_1} and~$(b)$ follows from definition of $\hat{h}.$

 After replace the value of $C_1$ in~\eqref{eq:for_replacing_C_1} from the above equation we get for $\tau \leq \taustar$, 
        \begin{align}
        \htauone=&c_a\lambda \tau-p \beta c_a\lambda\frac{\tau^2}{2} +\theta \tau+\hat{h}-c_f \label{value_of_h(tau,1,1)_for_second_case}
    \end{align}
    After combining~\eqref{eqn:bellman_h0tau11_subcase_1} and~\eqref{value_of_h(tau,1,1)_for_second_case} we get \begin{align*}
         \htauone=\begin{cases}
            c_a\lambda \tau-p \beta c_a\lambda\frac{\tau^2}{2} +\theta \tau+\hat{h}-c_f  \text{ for } \tau \leq \taustar\\
            h(0,0,1) \text{ for } \tau > \taustar
            \end{cases} \end{align*}

            Recall that from~\eqref{eq:h_0tau_1_0_full_st} and~\eqref{eqn:bellman_h0tau11_subcase_1}, $h(0,\tau,1,0)=h(0,\tau,1,1)-c_a\lambda \tau$ for $\tau\leq\taustar$. Replacing the value of $\htauone$ from~\eqref{value_of_h(tau,1,1)_for_second_case}, we get, \begin{align}
             \htauzero=-p \beta c_a\lambda\frac{\tau^2}{2} +\theta \tau+\hat{h}-c_f \text{ for $\tau \leq \taustar$}\label{h(tau,1,0)below_taustar}.   
            \end{align} Moreover, from~\eqref{eq:h_0tau_1_0_full_st} we observe that $\htauzero=h(0,0,0)$ for $\tau>\taub$. In the following, we obtain the expression of $\htauzero$ for $\taustar<\tau\leq \taub$. For this we consider the following equation from~\eqref{eq:h_0tau_1_0_full_st}.  
            \begin{align}
        \htauzero=& \frac{C_h-\theta}{\beta}+\Lotau \stackrel{(a)}= \frac{C_h-\theta}{\beta}+ e^{\beta \tau}L_{\tau}(0) \label{eq:h(tau,1,0) before derivative_queue}
        \end{align} where (a) follows from change of variables. We obtain the  derivative of $\htauzero$ w.r.t. to $\tau$ using~\eqref{eq:derivative_queue} as follows: 
        \begin{align}
            \dot{h}(0,\tau,1,0)=\beta e^{\beta \tau}L_{\tau}(0)-\beta\left(p\htauone+(1-p)\htauzero\right) \label{eq:h(tau,1,0) after derivative_queue}
        \end{align}
        By replacing the value of $\htauone=h(0,0,1)$ for $\tau>\taustar$ from~\eqref{eqn:bellman_h0tau11_subcase_1} and $e^{\beta \tau}L_{\tau}(0)=\htauzero-\frac{C_h-\theta}{\beta}$ from~\eqref{eq:h(tau,1,0) before derivative_queue} we get,
         \begin{align}
            \dot{h}(0,\tau,1,0)=\theta -C_h-p \beta h(0,0,1)+p \beta \htauzero\label{eq:final_h_tau,1,0_case_3}
        \end{align}
         By solving the above differential equation we get, \begin{align*}
        \htauzero=C_2e^{p \beta \tau}+\frac{C_h-\theta+p \beta h(0,0,1)}{p \beta}, 
    \end{align*}
    where $C_2$ is an integration constant. 
    To find the value of $C_2$, we use the value of $\htauzero$ at $\tau=\taub$ from~\eqref{eq:h_0tau_1_0_full_st},
    \begin{align*}
     h(\taub,1,0)=h(0,0,0)=&C_2e^{p \beta \taub}+\frac{C_h}{p \beta}+h(0,0,1)-\frac{\theta}{p \beta} \\
        \implies C_2=&-\frac{C_h}{p \beta}e^{-p \beta \taub}
        \end{align*}
        Since $C_h=0$, from~\eqref{eq:C_h_0}, we have 
        \begin{align*}
        \text{Hence, }\htauzero=h(0,0,0)\,\, \forall \taustar \leq \tau \leq \taub 
    \end{align*}
    By the definition of $\taub$, we have $\taub=\taustar$.
     \begin{align}
        \text{Hence, }\htauzero=h(0,0,0)\,\, \forall  \tau \geq \taub \label{final_value_of_h(tau,1,0)_in_case_2}
    \end{align}
    Since, $\taub=\taustar$, after combining~\eqref{h(tau,1,0)below_taustar} and~\eqref{final_value_of_h(tau,1,0)_in_case_2} we get,
    $\htauzero=\begin{cases}-p \beta c_a\lambda\frac{\tau^2}{2} +\theta \tau-C_h \tau+h(0,0,1)-c_f \text{ for $\tau \leq \taub$}\\
        h(0,0,0) \text{ for } \tau>\taub
        \end{cases}$

 \subsection{Proof of 
 Lemma~\ref{lemma:case_3_values_queue}}
 \label{appendix:proof_of_lemma_7}
We first obtain the value of $\theta$. At $\tau=\tautil$, 
from~\eqref{eqn:bellman_h0tau11_subcase_2},
\[c_a\lambda\tautil+h(0,0,0)=h(0,0,1)\]
By replacing $\hzero=-\frac{\theta}{p\beta}+\hzeroone$ from~\eqref{eqn:rel_hq00_and_hq01} in the above equation, we obtain 
\begin{equation}
    \theta=p\beta c_a\lambda\tautil \label{eq:theta_subcase_2_queue}
\end{equation}
        
        We  derive  the value of $h(0,\tau,1,1)$ for $\tau\leq \bar{\tau}$. Following similar steps while obtaining~\eqref{value_of_h(tau,1,1)_for_second_case}  we get 
    \begin{align}
        {h}(0,\tau,1,1)=c_a\lambda \tau-p \beta c_a\lambda\frac{\tau^2}{2} +\theta \tau+\hat{h}-c_f \label{equation_h_tau_1_0_without_value_of_theta}
        \end{align}
   Substituting the value of $\theta=p \beta c_a \tilde{\tau} \lambda $ from~\eqref{eq:theta_subcase_2_queue} we get,
        \begin{align}
          h(0,\tau,1,1)=&c_a\lambda\tau+p \beta c_a \lambda \left(\tilde{\tau} \tau-\frac{\tau^2}{2}\right)-C_h \tau+\hat{h}-c_f \nonumber
         \end{align} 
         Hence, combining with~\eqref{eqn:bellman_h0tau11_subcase_2} we obtain
          and obtain 
        $ h(0,\tau,1,1)=$  
        \begin{align}
             \begin{cases}
            c_a\lambda\tau+p \beta c_a \lambda \left(\tautil \tau-\frac{\tau^2}{2}\right)-C_h \tau+ \hat{h}-c_f \,\, \text{ for } \tau \leq \taub \\
            c_a\lambda\tau+h(0,0,0) \text{ for } \taub<\tau\leq\tautil\\
            h(0,0,1) \text{ for } \tau > \tautil. 
        \end{cases}\label{eq:h_tau_1_1_subcase_2}
          \end{align}

After following similar steps in obtaining~\eqref{eq:h_001_via_h000} we obtain the following from~\eqref{hQ01_under_subcase_2}.
\begin{equation}
    \hzeroone=\frac{c_w \Qbar (\Qbar+1)}{2p\beta}-\frac{\Qbar\theta}{p\beta}+\hat{h}. \label{eq:h_001_via_h000_subcase_2}
\end{equation}
To compute $\taub$, $\tautil$, and $\Qbar$ we obtain the following equations. 
At $\tau=\taub$, from~\eqref{eq:h_tau_1_1_subcase_2}, 
\begin{align}
    &c_a\lambda\taub+p \beta c_a \lambda \left(\tautil \taub-\frac{\taub^2}{2}\right)-C_h \taub+\hat{h}-c_f\nonumber\\
    & \hspace{2 in}= c_a\lambda\taub+h(0,0,0) \nonumber\\
     &\implies p \beta c_a \lambda \left(\tautil \taub-\frac{\taub^2}{2}\right)-C_h \taub+ \hat{h}-c_f\nonumber\\
    &\hspace{2 in}\stackrel{(a)}= -\frac{\theta}{p\beta}+h(0,0,1) \nonumber\\
    &\implies p \beta c_a \lambda \left(\tautil \taub-\frac{\taub^2}{2}\right)-C_h \taub+ c_f\nonumber\\
    &\hspace{1.5 in}\stackrel{(b)}= -\frac{\theta}{p\beta}+\frac{c_w \Qbar (\Qbar+1)}{2p\beta}-\frac{\Qbar\theta}{p\beta} \nonumber\\
    &{\implies} (\Qbar{+}1)c_a\lambda\tautil{+}p \beta c_a \lambda \left(\tautil \taub-\frac{\taub^2}{2}\right){-} \frac{c_w \Qbar (\Qbar+1)}{2p\beta}\nonumber\\
    &\hspace{2 in}-C_h\taub+c_f\stackrel{(c)}{=}0 \label{eq:1_with_3_param}
\end{align} 
The equality~$(a)$ is obtained by replacing \[h(0,0,0)=-\frac{\theta}{p\beta}+h(0,0,1)\] from~\eqref{eqn:rel_hq00_and_hq01}. The equality~$(b)$ is obtained by replacing $h(0,0,1)$ from~\eqref{eq:h_001_via_h000_subcase_2}. The equality~$(c)$ is obtained by replacing value of $\theta$ from~\eqref{eq:theta_subcase_2_queue}.

The equation~\eqref{eq:1_with_3_param} provides a relation between $\Qbar,\taub$ and $\tautil$. Since there are three unknown variables, we need two more equations to obtain $\Qbar,\taub$ and $\tautil$. We will obtain these equations in the following discussion.

  We observe from~\eqref{eq:h_0tau_1_0_full_st},\begin{align}h(0,0,0)=h(0,\taub,1,0)&\stackrel{(a)}= \frac{C_h-\theta}{\beta}+L_{0}(\taub)\nonumber\\
  &\stackrel{(b)}=\frac{C_h-\theta}{\beta}+e^{\beta \taub}L_{\taub}(0).\label{eq:h_0_0_vs_Lbar_tau}\end{align} The equality~$(a)$ is obtained from~\eqref{eq:h_0tau_1_0_full_st} by substituting $\tau=\taub$. The equality~$(b)$ is obtained by change of variables in the integration in~\eqref{eq:lotau_queue}.
  Recall that, 
  \begin{align*}
\lefteqn{L_0(\taub){\coloneqq}} \\
& \int_{0}^{\infty}\beta e^{-\beta t}\left(ph(0,t+\taub,1,1)+(1-p)h(0,t+\taub,1,0)\right)dt.
\end{align*} We substitute the value of $h(\tau,1,0)$  and $h(\tau,1,1)$ for $\tau\geq\taub$ from~\eqref{eq:h_0tau_1_0_full_st} and~\eqref{eqn:bellman_h0tau11_subcase_2} in the above equation. And then from~\eqref{eq:h_0_0_vs_Lbar_tau}, 
  \begin{align}
       &\hzero=\frac{C_h-\theta}{\beta}{+}e^{\beta \taub}\int_{\taub}^{\tautil}\beta e^{-\beta t}p(c_a\lambda t{+}\hzero)dt\nonumber\\
       &\quad + e^{\beta \taub}\int_{\tautil}^{\infty}\beta e^{-\beta t}p \hzeroone dt{+}(1{-}p)\hzero\nonumber\\
      & =\frac{C_h-\theta}{\beta} + \frac{p c_a\lambda}{\beta}\left((1+\beta \taub)-e^{\beta (\taub-\tautil)}(1+\beta \taub)\right)\nonumber\\
      &\quad + p \hzero(1-e^{\beta(\taub- \tautil)})+e^{\beta (\taub-\tautil)} p(\frac{\theta}{p \beta}+\hzero)\nonumber\\
      &\quad +(1-p)\hzero \nonumber
  \end{align}
 
         We substitute $\hzeroone=\frac{\theta}{p\beta}+\hzero$ from~\eqref{eqn:rel_hq00_and_hq01} and $\theta=\lambda p \beta c_a \tilde{\tau} $ from ~\eqref{eq:theta_subcase_2_queue} in the above equation and obtain the following.  
         \begin{align}
            &\hzero{=}\frac{C_h-\theta}{\beta}+\frac{p c_a\lambda}{\beta}(1+\beta \taub)-\frac{p c_a\lambda}{\beta}e^{\beta (\taub-\tautil)}\nonumber\\
            &\quad \quad +\hzero\nonumber\\
            &\Rightarrow p c_a \lambda+C_h-p c_a \lambda \left(\beta(\tautil-\taub)+e^{-\beta(\tautil-\taub)}\right)=0 \label{eq:2_with_3_param}
        \end{align}
By following similar steps to obtain~\eqref{eq:floor_taustar_qstar} the following can be derived.   \begin{equation}
            \Qbar=\left\lfloor\frac{p\beta c_a\lambda \tautil}{c_w}\right\rfloor. \label{eq:3_with_3_param}
        \end{equation}
        Hence, we can obtain $\taub$, $\tautil$ and $\Qbar$ by solving~\eqref{eq:1_with_3_param},~\eqref{eq:2_with_3_param} and~\eqref{eq:3_with_3_param}. 

         \paragraph*{\it Lower bound on $C_h$}
           We have \[\beta(\tautil-\tau)+e^{-\beta(\tautil-\tau)}=1+\frac{ C_h}{ p c_a \lambda}\]from~\eqref{eq:2_with_3_param}. For $\taub=\tautil$, $C_h=0$. Since $\beta(\tautil-\tau)+e^{-\beta(\tautil-\tau)}$ is a strictly increasing function of $\tautil-\taub$, for $\tautil>\taub$, $C_h>0$.
           \paragraph*{ Upper bound on $C_h$}
           We show that \[C_h\leq p \beta c_f-p c_a \lambda (1-e^{-\beta \tau^0}),\] where \[\tau^0=\frac{2p\beta c_f+c_w\hat{Q}(\hat{Q}+1)}{2p\beta c_a\lambda(\hat{Q}+1)}.\]
          If $\tautil\leq \tau^0$, which  implies $\tautil-\taub\leq\tau^{0}$. Hence, it immediately follows that   $\beta (\tautil-\tau)+e^{-\beta (\tautil-\taub)}\leq \beta \tau^0+e^{-\beta \tau^0}$. After replacing $\beta (\tautil-\tau)+e^{-\beta (\tautil-\taub)}=1+
           \frac{C_h}{c_a\lambda}$ from~\eqref{eq:2_with_3_param}, we get, \[1+
           \frac{C_h}{c_a\lambda}\leq \beta \tau^0+e^{-\beta \tau^0}.\] Hence, we obtain, $C_h\leq p \beta c_f-p c_a \lambda (1-e^{-\beta \tau^0})=I$ ( by definition of $I$ in Theorem~\ref{thm:fin_cap_wait}).
           Now, we will show that $\tautil\leq\tau^0$. For $\tau=\tautil$ from~\eqref{eqn:bellman_h0tau11_subcase_2} \[c_a\lambda \tautil+h(0,0,0)=h(0,0,1)\leq \hat{h}\]
           The last inequality follows from~\eqref{eq:h_0_0_1_full_st}. After substituting $h(0,0,0)=-\frac{\theta}{p\beta}+h(0,0,1)$ from~\eqref{eqn:rel_hq00_and_hq01} and $h(0,0,1)$ from~\eqref{eq:h_001_via_h000_subcase_2}, we obtain \[\tautil\leq \frac{2p\beta c_f+c_w\bar{Q}(\bar{Q}+1)}{2p\beta c_a\lambda(\bar{Q}+1)}\eqqcolon \tau^0\]
           %Note that,  we replace the value of $\tautil=\tau^0$ in~\eqref{eq:3_with_3_param}, \[\Qbar=\left\lfloor\frac{2p\beta c_f+ c_w\bar{Q}(\bar{Q}+1)}{2c_w(\bar{Q}+1)}\right\rfloor\] will be same as~\eqref{eq:theta_first_case}. Hence, at $C_h=I$, $\Qbar=\Qhat$. 
\subsection{Proof of Indexability}\label{proof:indexability_wait}
\remove{
We write passive sets for content $n$ for different holding costs $C_h$ using Theorem~\ref{thm:fin_cap_wait} and~\eqref{eq:extended_policy}.
From~\eqref{eq:extended_policy},
$\pi^{\ast}(s)=2$ for all $s\in\{(Q,\tau,1,0): Q>0,  \tau\geq 0\}$ for all $C_h\geq 0$. Furthermore, for $C_h=0$, from Theorem~\ref{thm:fin_cap_wait} and~\eqref{eq:extended_policy},
$\pi^{\ast}(s)=2$ for all $s\in\{(Q,\tau,1,1), (Q,0,1): Q<Q^{\ast}\}$. Hence, 
\begin{align*}
    {\cal P}^n(0)= &\left\{(Q,\tau,1,1),(Q,0,1): Q<Q^{\ast} \right\}\\
    &\cup\left\{(Q,\tau,1,0),(Q,0,0):Q>0,\tau\geq 0 \right\}.
\end{align*}
Following similar arguments, for $0<C_h<I$, 
\begin{align*}
\mathcal{P}^n(C_h)=&\left\{(0,\tau,1,0), (Q,0,1): \tau \geq \taub(C_h), Q\leq \bar{Q}(C_h)\right\} \\
& \cup\left\{(Q,\tau,1,0) ,(Q,0,0):Q>0,\tau\geq 0\right\}\\
&\cup\left\{(Q,\tau,1,1):Q\leq \bar{Q}(C_h),\tau\geq 0\right\}.
\end{align*} 
 where we write $\taub(C_h)$ and $\bar{Q}(C_h)$ to explicitly show the dependence of $\taub$ and $\bar{Q}$ on $C_h$. From Lemma~\ref{uniqueness_lemma_wait}, $\taub(C_h)$ and $\bar{Q}(C_h)$ are decreasing and nondecreasing functions of $C_h$, respectively. Hence ${\cal P}^n(C_h)$ is nondecreasing set for $0\leq C_h\leq I$. Finally, for $C_h>I$, $\mathcal{P}(C_h)=\bar{\cal S}$, and hence, $\mathcal{P}(\infty)=\bar{\cal S}$. 

Clearly, $\mathcal{P}^n(C_h)$ is nondecreasing for all $C_h \geq 0$.
Hence, content $n$ is indexable.}

We characterize the passive sets for content $n$ for different holding costs $C_h$ using Theorem~\ref{thm:fin_cap_wait} and~\eqref{eq:extended_policy}. For clarity, we omit the content index $n$ from the notation. To prove indexability, we will show that $\mathcal{P}(C_h)$ is nondecreasing for all $C_h \geq 0$~\cite{ayesta2021computation}.

First, define the set of states that are passive for every $C_h\ge0$:
\[
\mathcal F
:=
\{(Q,\tau,1,0), Q>0,\ \tau\ge0\}
\cup
\{(Q,0,0), Q\ge0\}.
\]
From~\eqref{eq:extended_policy},
$\pi^{\ast}(s)=2$ for all $s\in\{(Q,\tau,1,0): Q>0,  \tau\geq 0\}$ for all $C_h\geq 0$. We assign the passive dummy action $2$ to states $(Q,0,0)$,
$Q\ge0$, where no decision is taken. Hence, $\mathcal{F}\subseteq \mathcal{P}(C_h)$ for $C_h\geq 0$. 
For a queue threshold $q$, define
\begin{align}
\mathcal G(q)\coloneqq \{(Q,0,1),0\le Q<q\}
\cup\nonumber\\
\{(Q,\tau,1,1),0<Q<q,\ \tau\ge0\}.\label{def:g_q}
\end{align}
The cached-requested states $(0,\tau,1,1)$ and $(0,\tau,1,0)$ are treated separately, since their optimal actions depend on the age thresholds.

We consider three cases based on the values of $C_h$. 
\paragraph{$C_h=0$}From Theorem~\ref{thm:fin_cap_wait} and~\eqref{eq:extended_policy},
$\pi^{\ast}(s)=2$ for all $s\in\mathcal{G}(\Qstar)$. For $s\in\{(0,\tau,1,1),(0,\tau,1,0),\tau\geq 0\}$, the optimal actions $\in\{0,1\}$ if $\Qstar=0$ (from Theorem~\ref{thm:fin_cap_wait}). For $s\in\{(0,\tau,1,1),\tau> \taustar,\Qstar>0\}, \pi^\ast(s)=2$. Hence,  
\begin{align}
    \mathcal{P}(0)= \mathcal{F}\cup\mathcal{G}(\Qstar)\cup \{(0,\tau,1,1),\tau> \taustar,\Qstar>0\}.\label{def:P_0}
\end{align}
\paragraph{$0<C_h\le I$} 
From Theorem~\ref{thm:fin_cap_wait} and~\eqref{eq:extended_policy},
$\pi^{\ast}(s)=2$ for all $s\in\mathcal{G}(\Qbar(C_h))$. Moreover, for $s\in\{(0,\tau,1,0),\tau> \bar{\tau}(C_h)\}$, $\pi^\ast(s)=2$ from Theorem~\ref{thm:fin_cap_wait}. We define the following set \[\mathcal{R}(C_h)\coloneqq\begin{cases}
    \{(0,\tau,1,1),\bar{\tau}(C_h)<\tau\le \tilde{\tau}(C_h)\} \text{ if }\bar{Q}(C_h)=0,\\
    \{(0,\tau,1,1),\tau>\bar{\tau}(C_h)\} \text{ if }\bar{Q}(C_h)>0,
\end{cases}\]
 where we write $\taub(C_h)$ and $\bar{Q}(C_h)$ to explicitly show the dependence of $\taub$ and $\bar{Q}$ on $C_h$.
For every $s\in\mathcal{R}(C_h)$ the optimal actions from Theorem~\ref{thm:fin_cap_wait} is \[\pi^\ast(s)\begin{cases}
   = 4 \text{ if }\bar{Q}(C_h)=0,\\
    \in\{2,4\} \text{ if } \bar{Q}(C_h)>0.
\end{cases}\]
Hence, we have \[\mathcal{P}(C_h)=\left\{(0,\tau,1,0),\tau >\taub(C_h)\right\} \cup\mathcal{R}(C_h)\cup\mathcal{G}(\bar{Q}(C_h))\cup \mathcal{F}.\]
We first show that $\mathcal{P}(0)\subseteq \mathcal{P}(C_h)$ for any $0<C_h\leq I$. Note that, if $\Qbar(C_h)=0$ then $\Qstar=0$ from Lemma~\ref{uniqueness_lemma_wait}. Then ${\cal G} (\Qstar)={\cal G} (\Qbar(C_h))=\phi$. Hence, from~\eqref{def:P_0} $\mathcal{P}(0)=\mathcal{F}\subseteq\mathcal{P}(C_h)$. If $\Qbar(C_h)>0$ but $\Qstar=0$ then also the monotonicity of the passive set is preserved by the above argument, i.e., ${\cal P}(0)\subseteq {\cal P} (C_h)$. We remain to show the monotonicity for the case $\Qbar(C_h)>0$ and $\Qstar>0$. Since, $\Qstar\le \Qbar(C_h)$, ${\cal G}(\Qstar)\subseteq {\cal G} (\Qbar(C_h))$ by definition~\eqref{def:g_q}.  Furthermore, since $\taub(C_h)\leq\taustar$ from Lemma~\ref{uniqueness_lemma_wait}, we have $\{(0,\tau,1,1),\tau>\taustar\}\subseteq \{(0,\tau,1,1),\tau>\taub(C_h)\}$. Hence, ${\cal P}(0)\subseteq {\cal P}(C_h)$ for $0<C_h\leq I$.

Now we show that ${\cal P}(C_h)\subseteq {\cal P}(C_h')$ if $0<C_h\leq C_h'\leq I$.   From Lemma~\ref{uniqueness_lemma_wait}, $\taub(C_h)$ and $\tautil(C_h)$ are decreasing and increasing functions of $C_h$, respectively. Hence, for $C_h\leq C_h'$, we have $\mathcal{R}(C_h)\subseteq\mathcal{R}(C_h')$ and \[\left\{(0,\tau,1,0),\tau > \taub(C_h)\right\}\subseteq \left\{(0,\tau,1,0),\tau > \taub(C_h')\right\}.\]Furthermore, $\Qbar(C_h)$ is a nondecreasing function of $C_h$, then by definition~\eqref{def:g_q} $\mathcal{G}(\Qbar(C_h))\subseteq \mathcal{G}(\Qbar(C_h'))$. Hence, $\mathcal{P}(C_h)$ is nondecreasing set for $0\leq C_h\leq I$. 
  \paragraph{$C_h>I$} For $s\in\{Q,0,1,Q\geq0\}$, the optimal action $\pi^\ast(s)\in\{2,3\}$ from Theorem~\ref{thm:fin_cap_wait}. For $s\in\{(0,\tau,1,1),(0,\tau,1,0),\tau\geq0\}$, the optimal action is to evict the content (see the discussion in the proof of Theorem~\ref{thm:fin_cap_wait}). For $Q>0$ and every $\tau \ge 0$~\eqref{eq:extended_policy} gives
$\pi^\ast((Q,\tau,1,1))=\pi^\ast((Q,0,1))\in\{2,3\}$.
The remaining states belong to $\mathcal F$. Hence, we have $\mathcal{P}(C_h)=\bar{\mathcal{S}}$, and hence, $\mathcal{P}(\infty)=\bar{\mathcal{S}}$.

Clearly, $\mathcal{P}(C_h)$ is nondecreasing for all $C_h \geq 0$.
Hence, content $n$ is indexable.
  \subsection{Proof of Lemma~\ref{uniqueness_lemma_wait}}
  \label{apndx:prf_uniqueness_lemma}
 We need to show the following equations have unique solutions.
  \begingroup
     \allowdisplaybreaks
      \begin{align}
    & \beta pc_a \lambda (\tautil \taub{-}\frac{\taub^2}{2}){-}C_h \taub{+}(\bar{Q}+1)c_a \lambda \tautil {-}c_f{-}\frac{c_w\bar{Q}(\bar{Q}+1)}{2p\beta}=0\label{quad_eq} \\
     &\beta (\tautil-\taub)+e^{-\beta (\tautil-\taub)}-1-\frac{C_h}{ p c_a \lambda}=0\label{exponential_eq} \\
     &\bar{Q}=\left\lfloor \frac{p\beta c_a\lambda\tautil}{c_w}\right\rfloor \label{floor_eq}
\end{align}
\endgroup
Let $x\coloneqq \beta(\tautil-\taub)$. By definition of $\taub$ and $\tautil$, $x\ge 0$, since. Hence, we can rewrite~\eqref{exponential_eq} as $x+e^{-x}=1+\frac{C_h}{ p c_a \lambda}$.
         We note that $x+e^{-x}$ is an increasing monotone function of $x$. Hence~\eqref{exponential_eq} has a unique solution. Furthermore,  whenever $C_h=0$, $x=0$ and $x>0$ for $C_h>0$, and as $C_h$ increases $x$ also increases. Let the solution of $x+e^{-x}=1+\frac{C_h}{ p c_a \lambda}$ be $\fng$, where $g$ is some function. Although, it is a function of $c_a,C_h,\lambda,\text{ and }p$ . Since $c_a, \lambda$ and $p$ are fixed and $C_h$ can be varied, with abuse of notation we use $\fng$,   Hence, $\beta(\tautil-\taub)=\fng \implies \tautil=\frac{\fng}{\beta}+\taub$. Consider~\eqref{quad_eq} at $\taub$ and replace the value of $\tautil$, and we get,
        \begin{align}
                &c_a \lambda p \beta \left(\frac{\fng}{\beta}\taub+\frac{\taub^2}{2}\right)\nonumber-C_h \taub+(\Qbar+1)c_a \lambda(\frac{\fng}{\beta}+\taub)\nonumber\\
                &\hspace{1.65 in}-\frac{c_w\bar{Q}(\bar{Q}+1)}{2p\beta}-c_f=0 \nonumber\\     &\taub^2+2\taub(\frac{\fng}{\beta}-\frac{C_h}{c_a\lambda p\beta}+\frac{\Qbar+1}{p\beta})\nonumber\\
    &\quad-2\left(\frac{c_f}{c_a\lambda p \beta}{-}\frac{(\Qbar+1)\fng}{p\beta^2}+\frac{c_w\bar{Q}(\bar{Q}+1)}{2c_a\lambda p^2\beta^2}\right)=0\nonumber
           \end{align}
          Let $f(\Qbar)\coloneqq \frac{\fng}{\beta}-\frac{C_h}{c_a\lambda p\beta}+\frac{\Qbar+1}{p\beta}$ Hence,
           \begin{align}
             \taub&=-f(\Qbar)\nonumber\\
               & +\sqrt{(f(\Qbar))^2+2\left(\frac{c_f}{c_a\lambda p \beta}-\frac{(\Qbar+1)\fng}{p\beta^2}+\frac{c_w\bar{Q}(\bar{Q}+1)}{2c_a\lambda p^2\beta^2}\right)}.\label{eq:taub_in_terms_Qbar}
           \end{align}
           Similar to Lemma~\ref{lemma:uniqueness_of_qstar_and_taustar}, $\Qbar$ and $\taub$ are unique solutions to~\eqref{eq:taub_in_terms_Qbar} and~\eqref{floor_eq}. Since $\tautil=\frac{\fng}{\beta}+\taub$,  $\Qbar$, $\taub$ and $\tautil$ are unique solutions to~\eqref{quad_eq},~\eqref{exponential_eq} and~\eqref{floor_eq}.

Since $\fng+e^{-\fng}=1+ \frac{C_h}{pc_a\lambda}$ or $\fng-\frac{C_h}{pc_a\lambda}=(1-e^{-\fng})$. We make the following observations: \begin{enumerate}[(a)]
           \item Since $\fng\geq0$ we observe from the above equation that $\fng - \frac{C_h}{pc_a\lambda}\geq 0$ or $\frac{\fng}{\beta}-\frac{C_h}{c_a\lambda p\beta}\geq 0$.
           \item Since $\fng$ increases as $C_h$ increases, $\fng-\frac{C_h}{pc_a\lambda}$ increases as $C_h$ increases, $f(\Qbar)$ increases. 
           \end{enumerate}

Following these facts $\taub$ is a decreasing function of $C_h$. After taking the derivative of $\tautil$ w.r.t. $C_h$, we can show that, $\tautil$ is am increasing function of $C_h$. Since $\Qbar$ is proportional to $\tautil$ from~\eqref{floor_eq}, $\Qbar$ is also increasing in $C_h$. The lower bounds and upper bounds of $\taub$ and $\tautil$ can be obtained from~\eqref{exponential_eq} by equating $\taub=\tautil$ and $\taub=0$. The lower bound and upper bound of $\Qbar$ can be obtained by using the lower and upper bounds of $\tautil$, respectively.  

  \remove{        
  The first part of proof of Lemma~\ref{uniqueness_lemma_wait} follows from Lemma~\cite[Lemma 3]{koley2024fresh}. 
  Since $\tautil$ increasing in $C_h$, from~\eqref{eq:3_for_uniqueness_wait} $\theta$ is a nondecreasing function of $C_h$. }
  
  \subsection{Connection with Optimal Dispatching of a Poisson Process~\cite{Ross_1969}}

  The following problem is considered in~\cite[Section 5]{Ross_1969}. \remove{ Items arrive at a processing plant at a Poisson rate $\beta$. The controller needs to dispatch these items. The controller can dispatch any number of items. Each dispatch is assumed to incur a cost $c_f$, and there is a per unit waiting cost $c_w$ for each items.  Let $a(t)$ denotes the action taken by the controller at $t$;  $a(t) \in \mathcal{A}:=\{0,1\}$, where $a(t)=0$ denotes {\it wait}, $a(t)=1$ denotes {\it dispatch}.
   
   The goal is to minimize the time averaged cost as follows.
  \begin{equation}
      \min_{\pi\in\Pi} \lim_{T \to \infty} \frac{1}{T} \mathbb{E}\left[ c_f \cdot D(T) + \int_0^T c_w Q(t) \, dt \right]
  \end{equation}
  where $D(T)$ is the total number of dispatches until time $T$ and $Q(t)$ denotes the number of items not dispatched at time $t$ and $\Pi$ is a set of stationary policies. This problem can be formulated as Markov decision process considering $(Q(t),\tau(t))$ and $a(t)$ as state and action, where $\tau(t)$ is time elapsed since last dispatch.  
  
 Assume $\bar{\pi}$ is an arbitrary stationary policy. By changing the policy $\bar{\pi}$, such that each action $\bar{\pi}(Q(t),\tau(t)) = 1$ will dispatch
at the time of arrival of the last item instead of time $t$, we come up with a new policy $\pi'$
 which is a modified version of $\bar{\pi}$. Such
policy $\pi'$
 will not perform worse than $\bar{pi}$ for the average
cost problem. Please refer to~\cite{10452408} for rigorous proof. Hence, we consider those policies whose decision epochs are  at the time of arrivals. 
}
 Let $t_k$ denote the arrival epoch of $k^{th}$ arrival. Let $Q_k$ and $\tau_k$ denote the queue length at $k$th epoch and time elapsed since the last dispatch. Let $a_k$ denote the action taken at epoch $k$. $a_k \in \{0,1\}$ where 0 denotes wait and 1 denotes dispatch.

Let $C_k$ denote the cost incurred at epoch $k$.

\[
C_k =
\begin{cases} 
c_w(Q_k + 1)(t_{k+1} - t_k) & \text{if } a_k = 0, \\
c_f & \text{if } a_k = 1.
\end{cases}
\]

We can pose the minimization problem as

\[
\min \lim\limits_{T \to \infty} \frac{1}{T} \mathbb{E} \left[ \sum_{k=1}^{A(T)} C_k \right]
\]

We can formulate the problem as MDP with state space ${\cal S} = \{ (Q, \tau) \mid Q \in \mathbb{Z}_+, \tau \in \mathbb{R}_+ \}$. 

Given $(Q_k,\tau_k)=(Q,\tau)$ the state in the next epoch will be as follows. 
\[
(Q_{k+1}, \tau_{k+1}) =
\begin{cases} 
(Q_k + 1, \tau_k + \Delta \tau) & \text{if } a_k = 0 \\
(0, \Delta \tau) & \text{if } a_k = 1
\end{cases}
\]

where $\Delta \tau \sim \text{Exponential}(\beta)$.
We can write the Bellman equation from~\cite{bertsekas2011dynamic} as follows.

\begin{align}
&h(Q, \tau)\nonumber\\
=& \min \left\{ -\frac{\theta}{\beta} + \frac{c_w(Q+1)}{\beta} + \int_{0}^{\infty} \beta e^{-\beta t} h(Q+1, t+\tau) dt, \right.\nonumber\\
 &\left.\quad-\frac{\theta}{\beta} +c_f+ \int_{0}^{\infty} \beta e^{-\beta t} h(0, t) dt \right\}\label{bellman_eqn_with_queue_infinite_capacity_ross}
\end{align}
where $h$ is the relative cost function and $\theta$ is the optimal average cost. 

\begin{lemma} \label{lemma:opt_policy}
    There following policy $\pi$ is optimal \[
\pi(Q, \tau) =
\begin{cases} 
0 & \text{for } Q < \Qstar \\
1 & \text{for } Q \geq \Qstar
\end{cases}
\]
where $\Qstar$ and $\theta$ are the solutions to the following equations. 
    \begin{align}
    \frac{c_w\Qstar(\Qstar+1)}{2\beta}-\frac{(\Qstar+1)\theta}{\beta}+c_f&=0.\label{eq:first_eq_ross}\\
     \Qstar&=\left\lfloor \frac{\theta}{c_w} \right\rfloor. \label{eq:2nd_eq_ross}
    \end{align}
\end{lemma}
\begin{IEEEproof}
    The proof is same as Lemma~\ref{lemma:threshold_policy}
\end{IEEEproof}
\paragraph{Solution to~\eqref{eq:first_eq_ross} and~\eqref{eq:2nd_eq_ross}}
Let $\Qbar$ and $\bar{\theta}$ be the solution to the following equations:
\begin{align}
    \frac{c_wQ(Q+1)}{2\beta}-\frac{(Q+1)\theta}{\beta}+c_f=0\label{eq:qbar}\\
     Q= \frac{\theta}{c_w}.\label{eq:theta}
\end{align}
\begin{align} \text{ Hence, } \bar{\theta}=\frac{2\beta c_f+c_w\Qbar(\Qbar+1)}{2(\Qbar+1)}=\frac{2\beta c_f+\bar{\theta}(\frac{\bar{\theta}}{c_w}+1)}{2(\frac{\bar{\theta}}{c_w}+1)}.
\label{eq:qbar_in_terms_of_qbar}
\end{align}
Set
\begin{align}
    \Qstar=\lfloor \Qbar\rfloor=\left\lfloor \frac{\bar{\theta}}{c_w} \right\rfloor. \label{eq:Qstar_in_terms_Qbar}
\end{align}
then obtain $\theta$ from~\eqref{eq:first_eq_ross} and hence, 
\begin{align}
    \theta=\frac{2\beta c_f+c_w\left\lfloor \frac{\bar{\theta}}{c_w} \right\rfloor(\left\lfloor \frac{\bar{\theta}}{c_w} \right\rfloor+1)}{2(\left\lfloor \frac{\bar{\theta}}{c_w} \right\rfloor+1)}
\end{align}
If we can show $\Qstar\leq \frac{\theta}{c_w}\leq \Qstar+1$, then $\Qstar$ and $\theta$ are solutions to~\eqref{eq:first_eq_ross} and~\eqref{eq:2nd_eq_ross}. In particular we have to show 
\begin{align}
    \left\lfloor\frac{\bar{\theta}}{c_w}\right\rfloor\stackrel{(a)}{\leq}\frac{2\beta c_f+c_w\left\lfloor \frac{\bar{\theta}}{c_w} \right\rfloor(\left\lfloor \frac{\bar{\theta}}{c_w} \right\rfloor+1)}{2c_w(\left\lfloor \frac{\bar{\theta}}{c_w} \right\rfloor+1)}\stackrel{(b)}{\leq} \left\lfloor\frac{\bar{\theta}}{c_w}\right\rfloor+1. \label{eq:verify}
\end{align}
\[\text{From~\eqref{eq:qbar_in_terms_of_qbar} } (\frac{\bar{\theta}}{c_w}+1)\bar{\theta}=2\beta c_f\Rightarrow c_w(\frac{\bar{\theta}}{c_w}+1)\frac{\bar{\theta}}{c_w}=2\beta c_f.\] 
\[\text{Hence }c_w(\left\lfloor\frac{\bar{\theta}}{c_w}\right\rfloor+1)\left\lfloor\frac{\bar{\theta}}{c_w}\right\rfloor\leq2\beta c_f.\]
This implies $(a)$ in~\eqref{eq:verify} holds. Again \[\text{From~\eqref{eq:qbar_in_terms_of_qbar} } (\frac{\bar{\theta}}{c_w}+1)\bar{\theta}=2\beta c_f\Rightarrow c_w(\frac{\bar{\theta}}{c_w}+1)\frac{\bar{\theta}}{c_w}=2\beta c_f.\]

\[\text{Hence }c_w(\left\lfloor\frac{\bar{\theta}}{c_w}\right\rfloor+1)
(\left\lfloor\frac{\bar{\theta}}{c_w}+1)\right\rfloor\geq2\beta c_f.\]

This implies $(b)$ in~\ref{eq:verify}. 
\paragraph{Value of $\Qstar$}
From~\eqref{eq:qbar},
\[\Qbar= \frac{1}{2}\left(\sqrt{{1+\frac{8\beta c_f}{c_w}}}-1\right)\]

%Note that, \[ \sqrt{\frac{2\beta c_f}{c_w}}-\frac{1}{2}\leq \frac{1}{2}\left(\sqrt{{1+\frac{8\beta c_f}{c_w}}}-1\right)\leq \sqrt{\frac{2\beta c_f}{c_w}}.\]
Recall from~\eqref{eq:Qstar_in_terms_Qbar}, 
\[\Qstar=\lfloor \Qbar\rfloor \]
Note that the decision epochs are request epochs the number of requests in the system when the fetching decision is taken is $\Qstar+1$. The optimal policy in~\cite[Section: Multiple Dispatches]{Ross_1969} suggests to fetch whenever the present number of the requests at the system is one of the two integer adjacent to $\sqrt{\frac{2\beta c_f}{c_w}}$.  
In the following will show $\Qstar+1$ will be one of the two integer adjacent to $\sqrt{\frac{2\beta c_f}{c_w}}$. 

Let $x=\sqrt{\frac{2\beta c_f}{c_w}}$ and $y=\frac{1}{2}\left(\sqrt{{1+\frac{8\beta c_f}{c_w}}}-1\right)=\frac{1}{2}\left(\sqrt{1+4x^2}-1\right)$. Note that, $(x-\frac{1}{2})\leq y\leq x$. We need to show $\lfloor x\rfloor \leq \lfloor y\rfloor+1\leq \lfloor x\rfloor+1$. There are different cases:
\begin{enumerate}
    \item $y<\lfloor x\rfloor$. Since $y>x-\frac{1}{2}$, $\lfloor y\rfloor+1=\lfloor x\rfloor$. 
    \item $\lfloor x\rfloor\leq y$. Since $y\leq x$, $\lfloor y\rfloor+1=\lfloor x\rfloor+1$. 
\end{enumerate}
Hence $\Qstar+1$ will be one of the two integer adjacent to $\sqrt{\frac{2\beta c_f}{c_w}}$.
\end{document}